\documentclass[useAMS,usenatbib,twocolumn]{mn2e}
\usepackage{natbib, graphics, epsfig}

\usepackage{times}
\usepackage{amsmath}
\usepackage{amssymb}
\usepackage{textcomp}
\newcommand{\sm}{\small}

\def\jcap{J. Cosmol.  Astropart. Phys.}

\def\apj{ApJ}

\def\apjl{ApJ}
\def\mnras{MNRAS}

\def\na{New Astronomy}

\def\nat{Nature}

\def\prd{Phys. Rev. D}

\title[Subhaloes in Self-Interacting Galactic Dark Matter Haloes]
      {Subhaloes in Self-Interacting Galactic Dark Matter Haloes}
      \author[M. Vogelsberger et al.] {\parbox{18.5cm}{
          Mark Vogelsberger$^{1}$\thanks{mvogelsberger@cfa.harvard.edu},
          Jesus Zavala$^{2,3}$\thanks{CITA National Fellow},
          Abraham Loeb$^{1}$
        }\vspace{0.3cm}\\
        $^1$Harvard-Smithsonian Center for Astrophysics, 60 Garden Street, Cambridge, MA 02138, USA\\
        $^2$Department of Physics and Astronomy, University of Waterloo, Waterloo, Ontario, N2L 3G1, Canada\\
        $^3$Perimeter Institute for Theoretical Physics, 31 Caroline St. N., Waterloo, ON, N2L 2Y5, Canada}

\begin{document}
\date{Accepted ???. Received ???; in original form ???}

\pagerange{\pageref{firstpage}--\pageref{lastpage}} \pubyear{2012}

\maketitle

\label{firstpage}

\begin{abstract}
We present N-body simulations of a new class of self-interacting
dark matter models, which do not violate any astrophysical constraints
due to a non-power-law velocity dependence of the transfer
cross section which is motivated by a Yukawa-like new gauge boson
interaction. Specifically, we focus on the formation of a Milky
Way-like dark matter halo taken from the Aquarius project and re-simulate it for a couple of representative cases in the
allowed parameter space of this new model.
We find that for these
cases, the main halo only develops a small core ($\sim1$~kpc) followed by a
density profile identical to that of the standard cold dark matter scenario 
outside of that radius.
Neither the subhalo mass function nor the radial number density of
subhaloes are altered in these models but there is a significant change
in the inner density structure of subhaloes resulting in the formation
of a large density core. As a consequence, the inner circular
velocity profiles of the most massive subhaloes differ significantly
from the cold dark matter predictions and we demonstrate that they are compatible
with the observational data of the brightest Milky Way dSphs in such a
velocity-dependent self-interacting dark matter scenario. Specifically, 
and contrary to the cold dark matter case, there are no subhaloes that are more concentrated 
than what is inferred from the kinematics of the Milky Way dSphs. We conclude
that these models offer an interesting alternative to the cold dark matter 
model that can reduce the recently reported tension between the
brightest Milky Way satellites and the dense subhaloes found in cold dark matter
simulations.
\end{abstract}


\section{Introduction}

The collisionless nature and low intrinsic thermal velocities of dark matter (DM) particles are two of the
fundamental hypotheses underlying the remarkably successful Cold Dark Matter (CDM) paradigm. The latter of
these hypotheses is related to a filtering mass scale that sets the minimum mass for self-bound DM structures (haloes)
to be typically of $\mathcal{O}$(1M$_{\oplus}$) and hence the CDM model predicts the existence of a large population of Earth-mass haloes 
in the local Universe. Although such prediction has yet to be unambiguously tested by observations, there is evidence of a potential tension
with the actual number of observed dwarf galaxies, both within the Milky-Way (MW) 
\citep[the so called ``missing satellite problem'', e.g.][]{Klypin1999,Moore1999}, and in the field
\citep[][as inferred from the HI velocity function measured by the ALFALFA survey]{Zavala2009,ALFALFA2011}.
This challenge to the CDM model is undoubtedly connected to the development of a successful theory of galaxy formation 
at the scale of dwarfs. For instance, it has been shown that the combination of astrophysical processes that inhibit star 
formation are able to solve the ``missing satellite'' problem \citep[e.g.][]{Koposov2009}, but the tension with dwarf 
galaxies in the field remains despite attempts to explain it without modifying the CDM model \citep{Ferrero2011}. 
It remains to be seen if galaxy formation models based on N-body simulations that seem to provide a good fit to the luminosity function at low masses \citep[][]{Guo2011}
are also successful in reproducing the HI observations of the velocity function.
Such a prevailing challenge strengthens the case for a solution based on a suppression of primordial small-scale density 
fluctuations. For example, models where DM is ``warm'', i.e., where DM particles have larger primordial thermal velocities, 
are able to alleviate the overabundance problem \citep[e.g.][]{Bode2001,Zavala2009} and at the same time, if their 
masses are $\gtrsim1$~keV, avoid current constraints on the matter power spectrum as measured by the Lyman$-\alpha$ 
forest \citep{Boyarsky2009}.

The second hypothesis, that of DM being essentially collisionless, has also been contested. 
The idea of self-interacting Dark Matter (SIDM) was first suggested by \cite{Carlson1992, Machacek1993, Laix1995} nearly 20 years ago.
More than a decade ago, \citet{SpergelSteinhardt2000} realised that the aforementioned overabundance problem of dwarf galaxies, and the 
observation of low surface brightness (LSB) galaxies having density cores \citep{deBlok1997} (which contradicts the high density cusps predicted by the CDM model \citep{NFW1996,NFW1997}) could be avoided if DM would be self-interacting. 
Current observations of LSB galaxies \citep{Kuzio2011} and MW dwarf spheroidals (dSphs) \citep{Walker2011} seem to confirm the presence 
of density cores in low-mass haloes. Since these galaxies are DM-dominated, it is challenging to invoke baryonic processes as the main mechanisms
responsible of altering so drastically the inner density profile of haloes. If DM is not cold, then haloes are
expected to have cores, although those associated with currently allowed WDM models seem to be too small to explain the observed cores of LSB galaxies 
\citep{Kuzio2011,VNDalal2011}. Haloes within WDM simulations, although less concentrated, seem to have similar profiles as 
their CDM counterparts and they do not show a clear sign of developing cores \citep[e.g.][]{Colin2000}. High resolution simulations within the
WDM cosmogony are however challenging \citep{Wang2007} and hence a consensus on the inner density profile of WDM haloes has not been reached.
SIDM models on the other hand, lead naturally to the development of a substantial core as was already shown by the first SIDM simulations 
\citep{Yoshida2000a,Yoshida2000b,Dave2001,Colin2002}. It remains to be seen if SIDM models are able to explain the observed cores of MW
dSphs and LSB galaxies. 

The first SIDM models assumed a constant scattering cross section and were quickly abandoned
since those that could solve the small-scale CDM problems seemed to violate several astrophysical constraints, such as the observed
ellipticity of the mass distribution of galaxy clusters \citep[e.g.][]{Miralda2002} and the survivability of satellite haloes \citep[e.g.][]{Gnedin2001}.
To avoid such constraints, simple ad hoc velocity-dependent cross sections of the form $1/v^\alpha$ were explored \citep[e.g.][]{Colin2002}, yielding
encouraging results that however lacked a proper underlying particle physics model. It has also been claimed that
these velocity-dependent SIDM models are not able to solve simultaneously the core problem in DM-dominated systems and the ``missing satellite
problem'' \citep[e.g.][]{DOnghiaBurkert2003}.

\citet{LoebWeiner2011} proposed that the possible existence of a Yukawa
potential among DM particles can resolve the challenges facing
SIDM with a constant cross section. The velocity dependence of scattering
through the massive mediator of this dark force (similar to a screened
Coulomb scattering in a plasma) could make scattering important at the low
velocity dispersion of dwarf galaxies but unimportant at the much higher
velocities encountered in galaxy clusters. The possibility of
exothermic reactions could in addition introduce a special velocity scale
around which the influence of the DM interaction peaks. The existence of
dark forces was studied earlier as a solution to cosmic ray anomalies
through enhanced dark matter annihilation \citep{ArkaniHamed2009}).

A recent analysis by \citet{Boylan2011a} puts forward an additional challenge to the CDM model. The authors
used simulated MW-like haloes in a CDM cosmology to show that the observed MW dSphs are 
only consistent with inhabiting relatively low-mass CDM subhaloes, leaving a population of more massive subhaloes with 
no galaxies associated to them, i.e., massive subhaloes of CDM MW-like haloes seem to be too dense to host the bright MW dSphs.
In a recent extension to this analysis, \citet{Boylan2011b} showed that this problem is unlikely to be solved invoking standard galaxy formation
processes based on CDM. This is one of the most serious
challenges faced by the CDM model and can perhaps be solved by invoking WDM \citep{Lovell2011} or, alternatively, also naturally avoided in
certain SIDM models as we explore in this work.

\begin{figure*}
\centering
\includegraphics[width=0.33\textwidth]{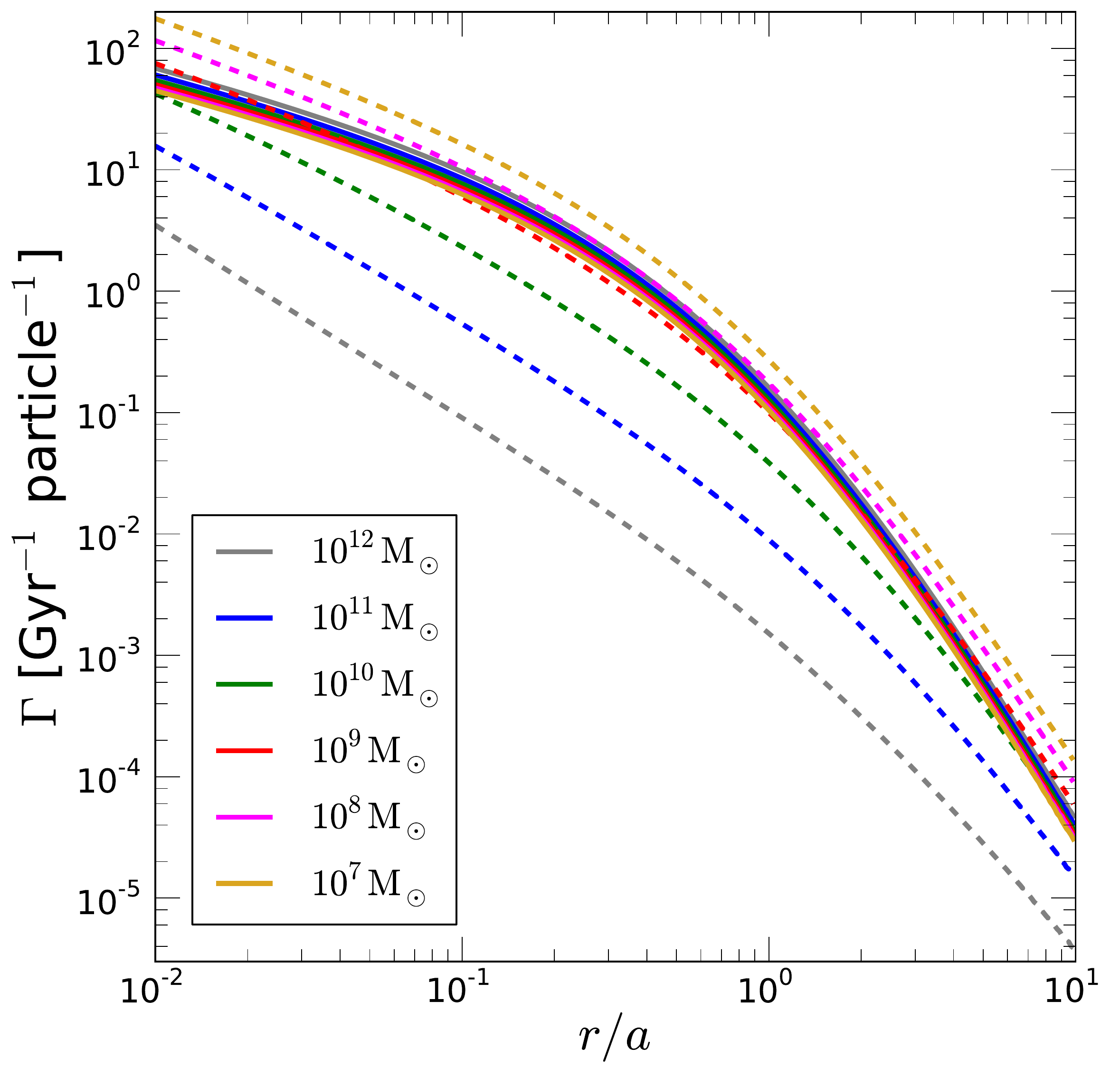}
\includegraphics[width=0.33\textwidth]{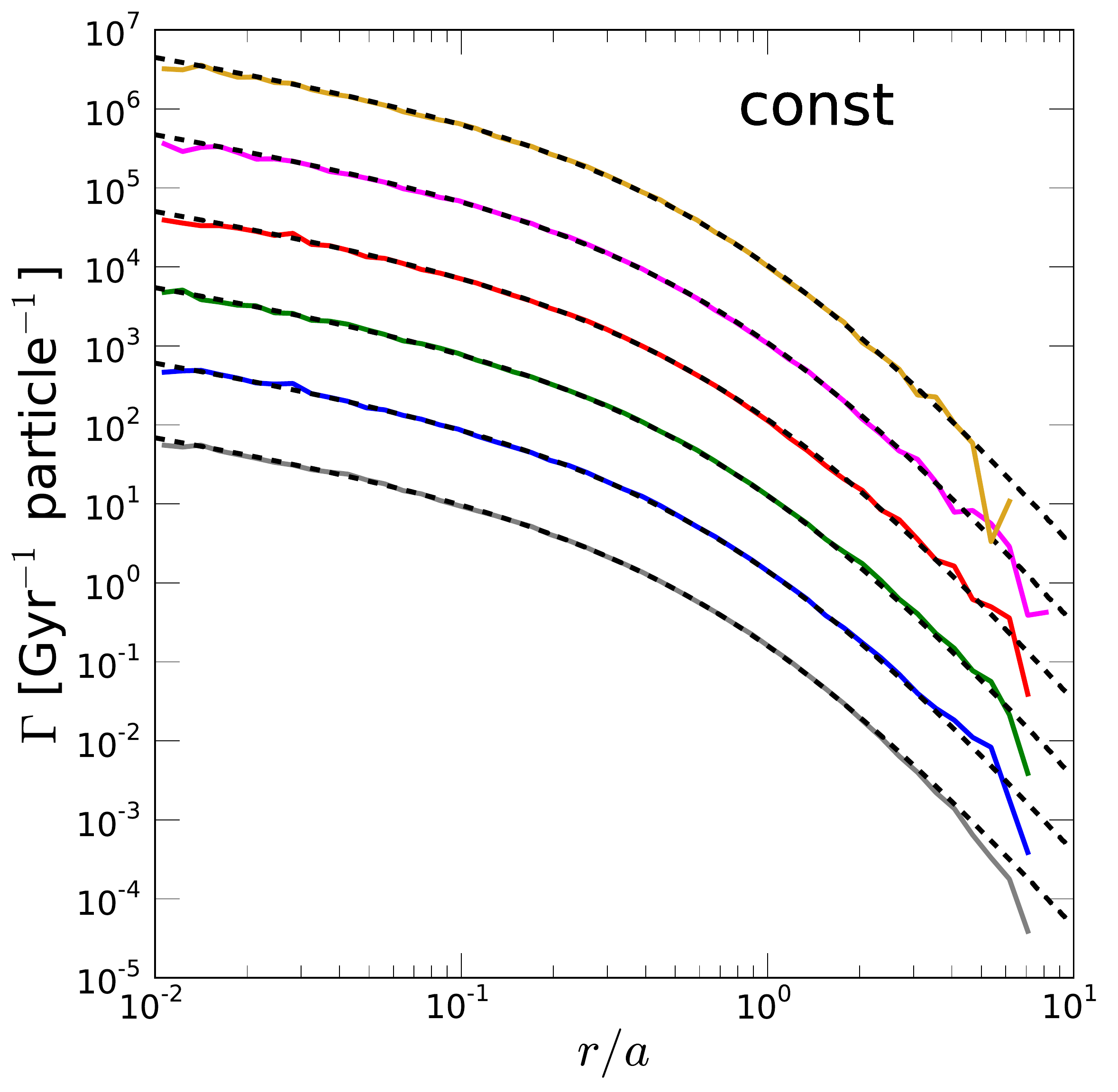}
\includegraphics[width=0.33\textwidth]{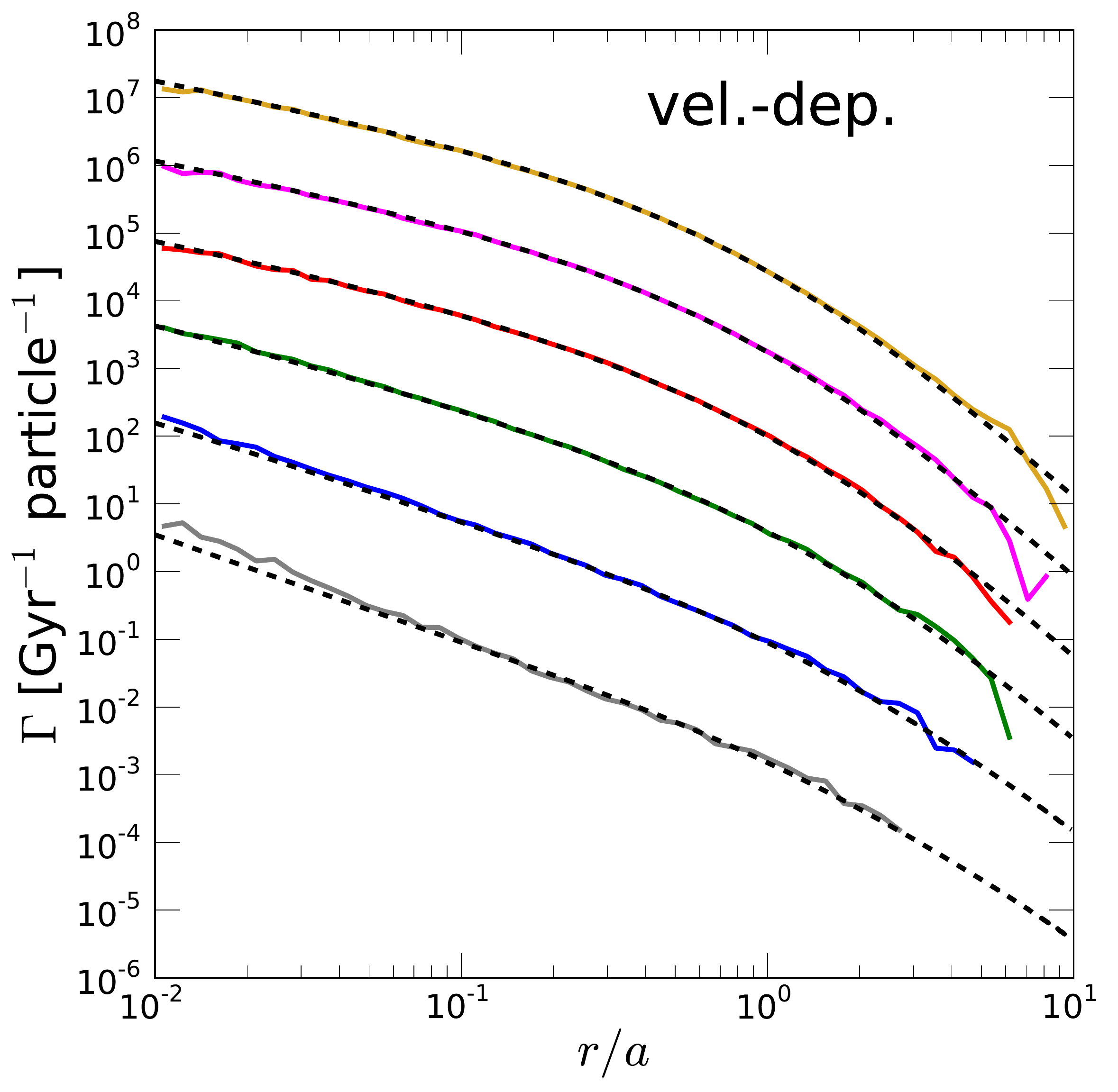}
\caption{Left panel: Analytic scatter rate profiles for Hernquist haloes of different masses following the average
cosmological concentration-mass relation. Dashed lines show the result for a velocity-dependent cross section ($v_{\rm max}=30~{\rm km}\,{\rm s}^{-1}$, $\sigma_T^{\rm max}/m_\chi=10~{\rm cm}^2\,{\rm g}^{-1}$), whereas
solid lines show the rates for a constant cross section ($\sigma_T^{\rm max}/m_\chi=10~{\rm cm}^2\,{\rm g}^{-1}$). The remaining two
panels show a comparison of these scatter rates with those obtained from a frozen N-body representation. The simulations were run
for $1~{\rm Gyr}$, and the different halo profiles are all offset by one dex for clarity. Solid lines show the N-body result, whereas black
dashed lines represent the analytic result repeated from the leftmost panel. Middle panel: constant cross section. Right panel:
velocity-dependent cross section.}
\label{fig:scatter_profiles_hernquist}
\end{figure*}

This is the first paper in a series whose objective is to study the properties of DM (sub)haloes within allowed velocity-dependent SIDM (vdSIDM) models
using state-of-the-art numerical simulations. In this work we only focus on the recently
discovered problem pointed out by \citet{Boylan2011a,Boylan2011b} and demonstrate how vdSIDM models can reduce the discrepancy between
observation and theoretical prediction. 
In Section~\ref{methods} we briefly describe the particle physics model we use
and present our numerical algorithm including test simulations. In Section~\ref{sims} we present simulations of a MW-like dark matter halo using the
initial conditions of one of the Aquarius haloes \citep{Springel2008} and compare its 
resulting subhalo population, for a couple of representative cases in the parameter space of the vdSIDM model, with the ones of the
standard collisionless CDM model. These models are then contrasted against observational data from the MW dSphs in
Section~\ref{dsphs}. Finally a summary and the conclusions of our work are given in Section~\ref{conclusion}.

\section{Methodology}\label{methods}

\subsection{Velocity-dependent SIDM (vdSIDM) models}\label{mod_sec}

We use a simplified particle physics model where the self-scattering between DM particles of mass $m_{\chi}$ is set by
an attractive Yukawa potential with coupling strength $\alpha_c$ mediated by a new gauge boson of mass $m_{\phi}$
(either scalar or vector) in the dark sector. We refer the reader to \cite{Feng2010b, Finkbeiner2011,LoebWeiner2011} for
details on the particle physics model and its available parameter space. If we only consider elastic interactions, 
the scattering problem is analogous to the screened Coulomb scattering in a plasma, 
which is well fitted by a transfer cross section given by:
\begin{equation}
\frac{\sigma_T}{\sigma_T^{\rm max}} \approx 
     \begin{cases}
       \frac{4\pi}{22.7}~\beta^2~{\rm ln}\left(1+\beta^{-1}\right),                    &\beta<0.1\\ \\
       \frac{8\pi}{22.7}~\beta^2~\left(1+1.5\beta^{1.65}\right)^{-1},                   &0.1<\beta<10^3\\ \\
       \frac{\pi}{22.7}~\left({\rm ln}\beta+1-\frac{1}{2}{\rm ln}^{-1}\beta\right)^2,  &\beta>10^3, 
     \end{cases}
\label{eq:cross}
\end{equation}
where $\beta=\pi v_{\rm max}^2/v^2=2\,\alpha_c\,m_{\phi}/(m_{\chi}v^2)$ and $\sigma_T^{\rm max}=22.7/m_{\phi}^2$, and $v$ is the
relative velocity of the DM particles. Here $v_{\rm max}$ is the velocity at which $(\sigma_T v)$ peaks at a transfer cross
section equal to $\sigma_T^{\rm max}$. 

The value of $\sigma_T^{\rm max}/m_{\chi}$ is constrained by different astrophysical measurements, the most stringent being: 
i) the observed ellipsoidal shape of haloes as implied by X-ray data of galaxy clusters and ellipticals \citep{Miralda2002,Feng2010a};
ii) avoidance of the gravothermal catastrophe leading to inner density profiles even steeper than those predicted in CDM \citep{Firmani2001};
and iii) avoidance of the destruction of subhaloes through collisions with high velocity particles from a larger parent halo 
\citep{Gnedin2001}. There is a summary of these and other constrains in Table I of \citet{Buckley2010} and in Figure~2 of 
\citet{LoebWeiner2011}: on the scales of dwarf galaxies, $\sigma_{\rm vel}\sim10\,{\rm km\,s^{-1}}$, 
the allowed values for the transfer cross section are roughly constrained from above by 
$\sigma_T^{\rm max}/m_{\chi}~\lesssim~ 35~{\rm cm^2\,{\rm g}^{-1}}$, and are much lower at  
$\sigma_{\rm vel}\sim100\,{\rm km\,s^{-1}}$, where the constraints are stronger by approximately two orders of magnitude. Since we are
interested in the possibility of producing cored density profiles for the haloes associated with the MW dSphs, we will take
two benchmark points in the $(\sigma_T^{\rm max}/m_{\chi},v_{\rm max})$ parameter space close to the aforementioned constraints that maximise
the self-interaction at the typical velocity dispersion of these dwarfs (see Table ~\ref{table:ref_points} in Section \ref{IC} below).

In this work we only consider elastic scattering leaving the cases of excited states and their associated exo- and endothermic 
interactions for a future analysis.

\subsection{Numerical technique}\label{tech}

To account for DM self-interactions we follow a standard Monte Carlo approach similar
to previous implementations \citep{KochanekWhite2000, Burkert2000, Yoshida2000a, Yoshida2000b, Dave2001, Craig2001, Colin2002, Donghia2003, KodaShapiro2011}, but
different from fluid smoothed particle hydrodynamics approaches as in \citet{Moore2000} and \citet{Yoshida2000a}.
 
We determine the scattering probability for every particle $i$ with each of its $k=38\pm5$
 nearest neighbours\footnote{This choice is to speed up the
neighbour search, but we checked that it does not affect any results.} $j$ in a time step $\Delta t_i$ by
\begin{equation}
  P_{ij}=\frac{m_i}{m_\chi} \, W(r_{ij},h_{i}) \, \sigma_T(v_{ij})v_{ij} \, \Delta t_i,
\label{eq:prob}
\end{equation}
where $m_i$ is the simulation particle mass, $v_{ij}$ is the relative velocity between particles $i$ and $j$, 
$\sigma_T/m_\chi$ is the scattering cross section per unit mass 
described in Section \ref{mod_sec}, $h_{i}$ the smoothing length enclosing the $k$ nearest neighbours of
particle $i$, and $W(r_{ij},h_{i})=w(r_{ij}/h_{i})$ is the cubic spline Kernel function in 3D normalisation:
\begin{equation}
w(q) =\frac{8}{\pi} \left\{
\begin{array}{ll}
1-6 q^2 + 6 q^3, &
0\le  q \le\frac{1}{2} ,\\
2\left(1-q\right)^3, & \frac{1}{2}< q \le 1 ,\\
0 , & q>1. 
\end{array}
\right.
\end{equation}

The time step $\Delta t_i$ is chosen small enough to avoid multiple scatterings during a time step by requiring that $\Delta t_i < \kappa~(\rho_i \, \sigma_T(\sigma_{{\rm vel},i})/m_\chi \, \sigma_{{\rm vel},i})^{-1}$, where 
$\sigma_{{\rm vel},i}$ is the local velocity dispersion at the position of particle $i$ calculated based on its $k$ neighbours, and we
set $\kappa=10^{-2}$, which is sufficiently small to avoid multiple scatterings during a step and usually smaller than the time step inferred from the dynamical time scale.
The total probability of a particle to interact with any of its neighbours is given by $P_i=\sum_j P_{ij}/2$, where the subscript
$j$ is limited to the $k$ neighbours. The factor $1/2$ accounts for the fact that a scatter event always involves two particles,
and we therefore need to divide by two to reproduce the correct scatter rate. We say that a collision takes place between particle $i$ and one of its $k$ nearest neighbours $j$ if
$x\leq P_i$, where $x$ is a uniformly distributed random number
in the interval $(0,1)$. To select the neighbour $j$ that is chosen for collision we sort them according to their distance
to particle $i$ and select the first neighbour $l$ that satisfies $x\leq \sum_i^l P_{ij}$. In the following 
we  assume that the self-interaction is isotropic. 

\begin{figure*}
\centering
\includegraphics[width=0.475\textwidth]{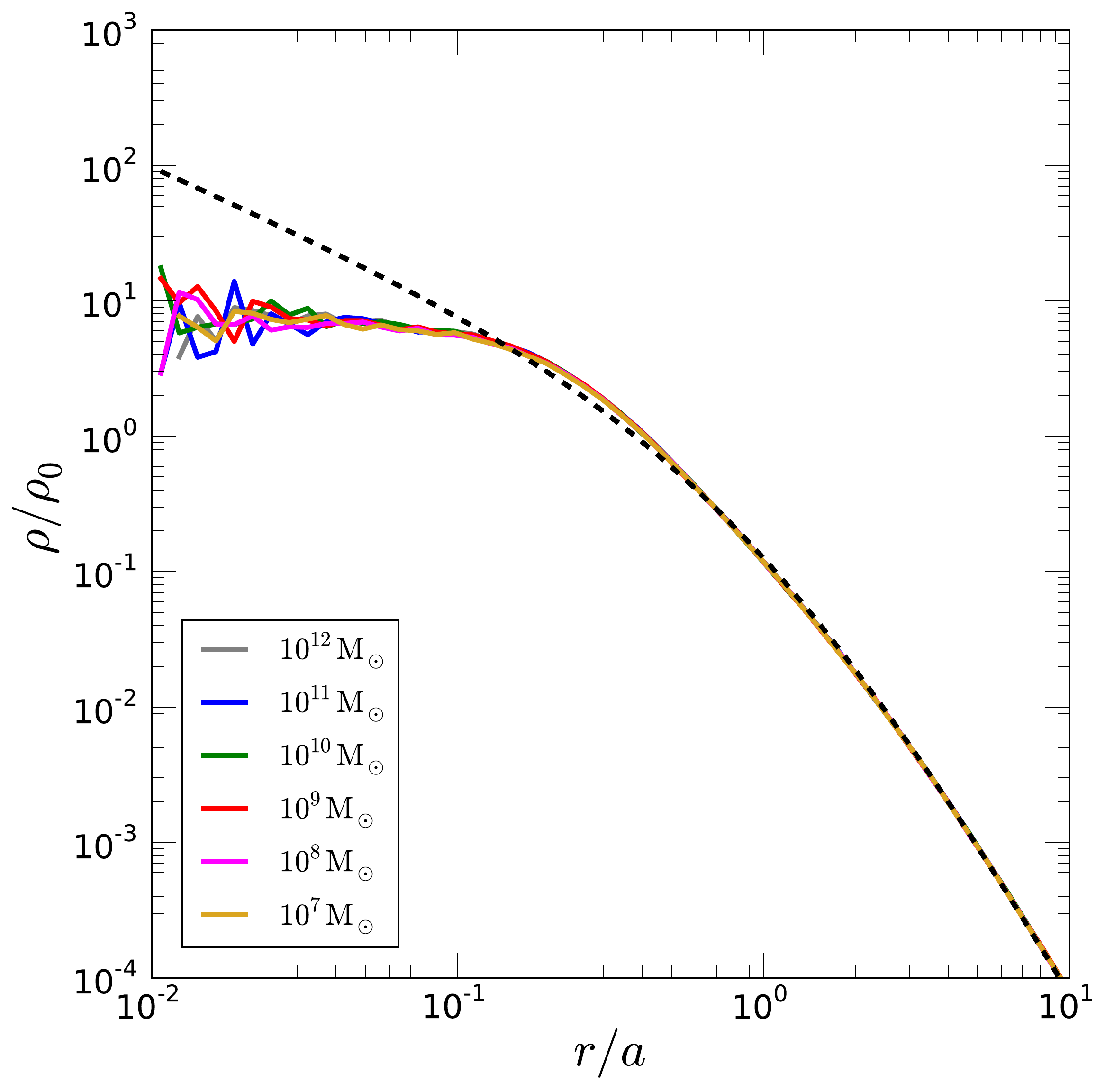}
\includegraphics[width=0.475\textwidth]{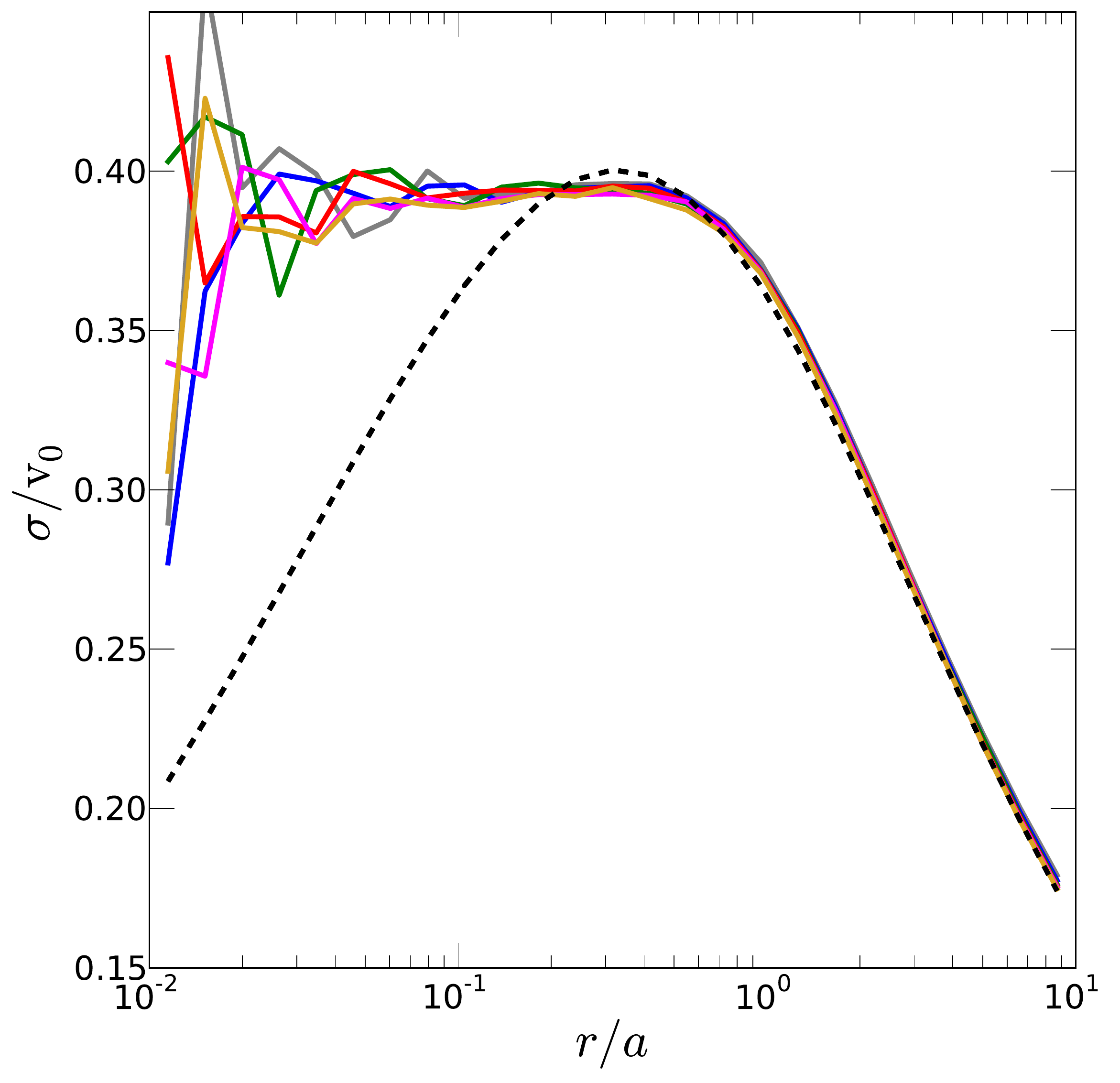}
\includegraphics[width=0.475\textwidth]{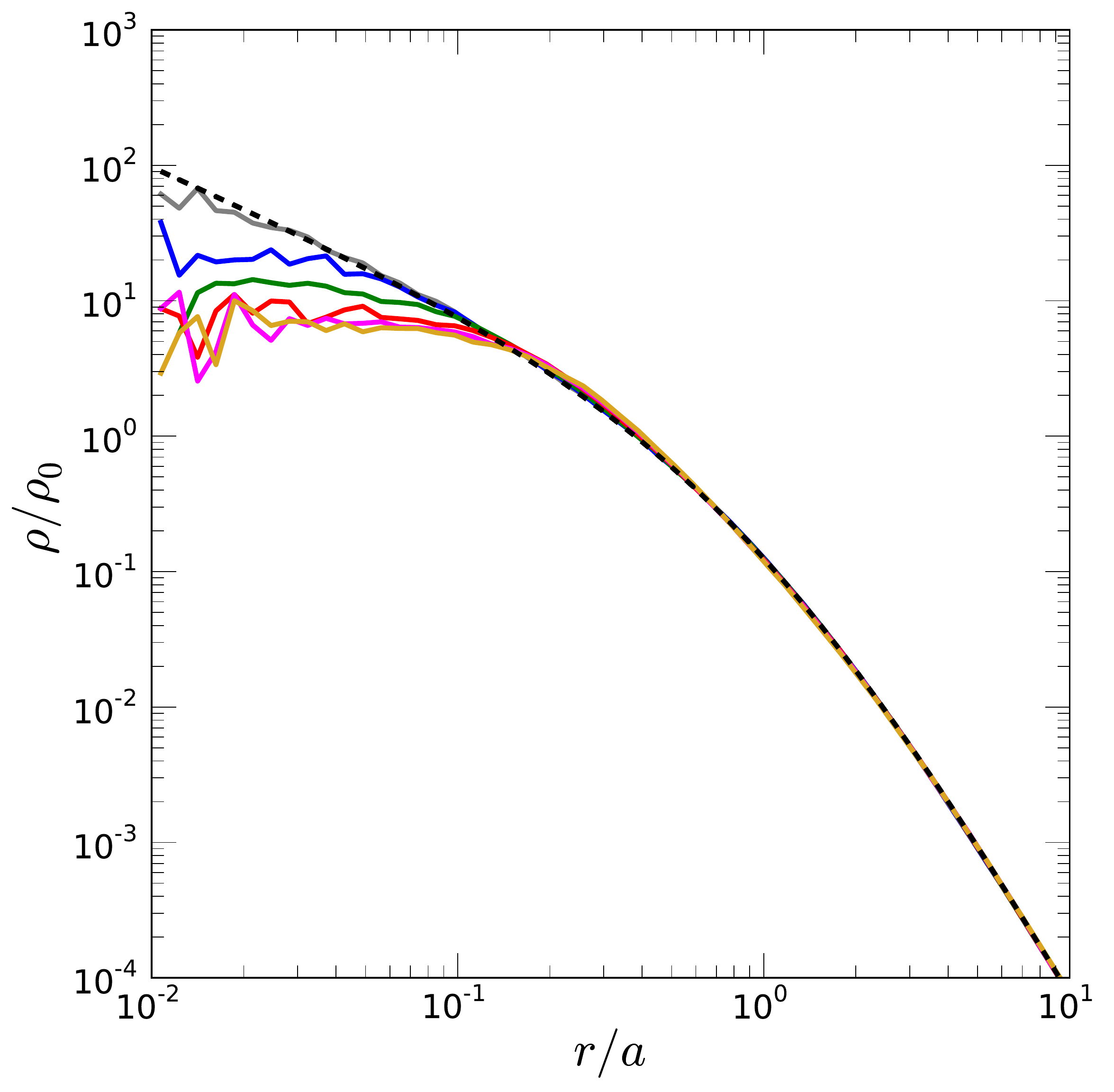}
\includegraphics[width=0.475\textwidth]{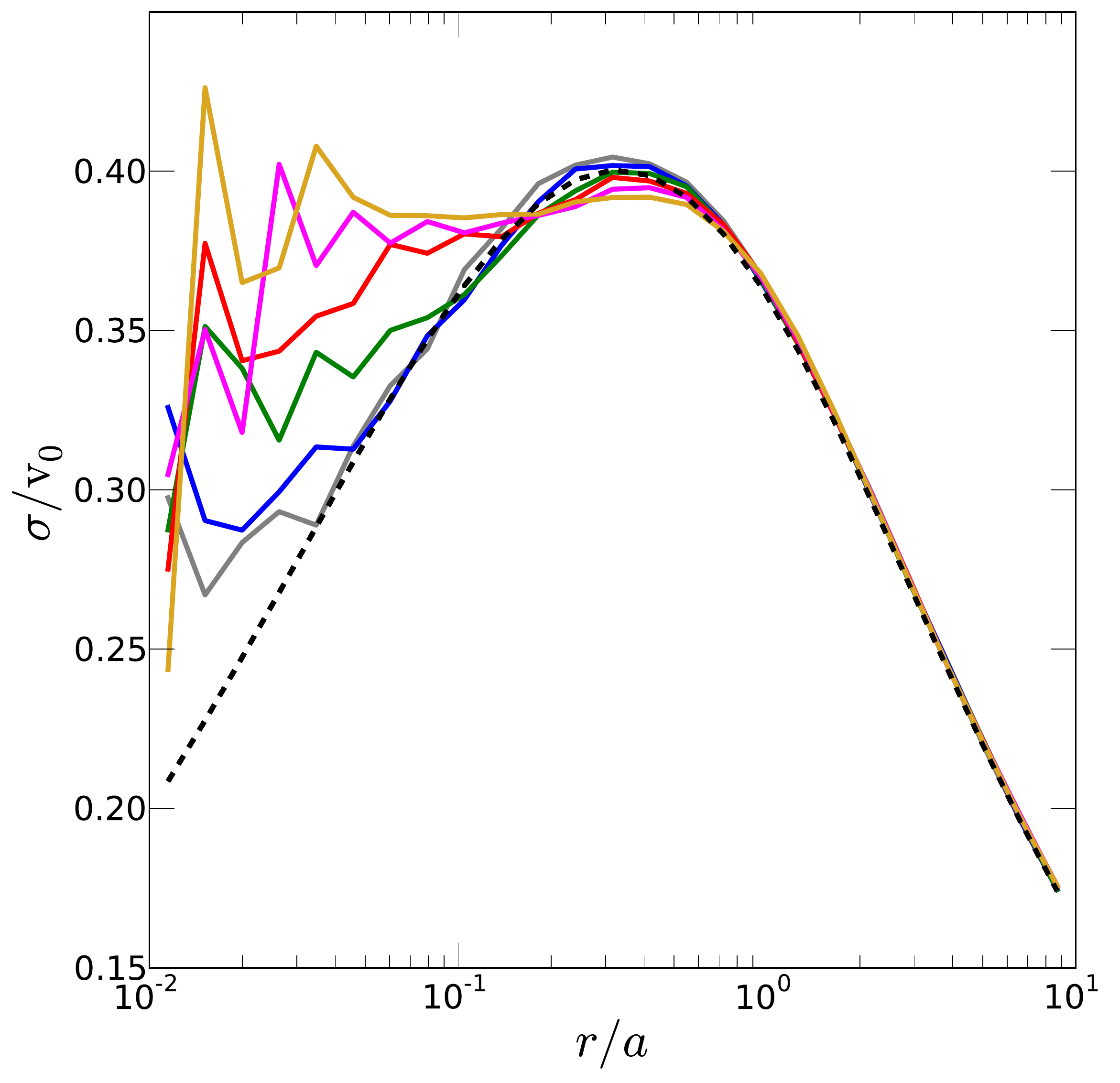}
\caption{Density (left panels) and velocity dispersion profiles (right panels) of haloes of different
masses. The top panels are for the case of a constant cross section ($\sigma_T^{\rm max}/m_\chi=10~{\rm cm}^2\,{\rm g}^{-1}$)
showing the profiles after $25~t_0$. Bottom panels are for the case of a velocity-dependent cross section
($v_{\rm max}=30~{\rm km}\,{\rm s}^{-1}$, $\sigma_T^{\rm max}/m_\chi=10~{\rm cm}^2\,{\rm g}^{-1}$) after $1~{\rm Gyr}$. In
scaled units, the constant cross section curves for all masses collapse to a single one. For the velocity-dependent case, evolution progresses faster
for lower mass systems, because $(\sigma_T v)$ peaks at a velocity of $30~{\rm km/s}$.}
\label{fig:scatter_profile_hernquist_live}
\end{figure*}

In the case of elastic scattering once a pair is tagged for collision we assign to each particle a new velocity given by:
\begin{align}
  \vec{v}_i&=\vec{v}_{cm}+(\vec{v}_{ij}/2)\,\hat{e},\nonumber\\
  \vec{v}_j&=\vec{v}_{cm}-(\vec{v}_{ij}/2)\,\hat{e},
\label{eq:vel}
\end{align}
where $\vec{v}_{cm}$ is the centre-of-mass velocity of the pair and $\hat{e}$ is a unit vector that we randomly draw from the
unit sphere. This procedure conserves energy and linear momentum, but not angular momentum. We have implemented this numerical scheme 
in {\sm GADGET-3} \citep[last described in][]{Springel2005}.

To test our implementation we apply it first to isolated haloes. For a region of volume $V$, the total number of scattering events is given by:
\begin{equation}
  \Gamma_{\rm tot}=\int_V\frac{\rho(\vec{x})^2}{2m_{\chi}^2}\left<\sigma_T v\right>(\vec{x})\,{\rm d}V,
\label{eq:int_rate}
\end{equation}
where $\rho(\vec{x})$ is the local DM density and $\left<\sigma_T v\right>(\vec{x})$ is the local thermal average of the
transfer cross section times the relative velocity. In the non-relativistic limit this is given by an average over a Maxwell-Boltzmann
distribution function:
\begin{equation}
  \left<\sigma_T v\right>(\vec{x})=\frac{1}{2\sigma_{\rm vel}^3(\vec{x})\sqrt{\pi}}\int (\sigma_T v) v^2e^{-v^2/4\sigma_{\rm vel}^2(\vec{x})}\,{\rm d}v,
\end{equation}
where $\sigma_{\rm vel}(\vec{x})$ is the local velocity dispersion. For given density and velocity distribution functions we can now calculate
the number of expected scattering events and compare this to the N-body / Monte Carlo results obtained with the technique presented in the paragraphs above.

As an example of the number of scattering events expected in a DM halo, we take a smooth spherical distribution of DM 
with a Hernquist density profile \citep[][]{Hernquist1990}:
\begin{equation}
  \rho(r)=\frac{Ma}{2\pi r}\frac{1}{(r+a)^3},
\label{eq:Hern}
\end{equation}
where $M$ is the total mass of the halo and $a$ its scale length. The velocity dispersion profile for the Hernquist halo follows from the
Jeans equation, which for an isotropic velocity distribution and using Eq.~(\ref{eq:Hern}) gives:
\begin{align}
  \sigma^2_{\rm vel}(r) & = \frac{GM}{12a}\left[\frac{12r(r+a)^3}{a^4}{\rm ln}\left(\frac{r+a}{r}\right) \right. \\
                      & - \left. \frac{r}{r+a}\left(25+52\left(\frac{r}{a}\right)+42\left(\frac{r}{a}\right)^2+12\left(\frac{r}{a}\right)^3\right)\right].\nonumber
\end{align}
It is then straightforward to compute the scattering rate using Eq.~(\ref{eq:int_rate}). To compare
these analytical expectations with N-body simulations, it is necessary to take into account the mass resolution of the simulation.
We therefore need to multiply  Eq.~(\ref{eq:int_rate}) with $m_{\chi}/m_{dm}$, where $m_{dm}$ is the DM particle mass of the simulation, which
yields the number of scatter events in the simulation volume.

The left panel of Figure~\ref{fig:scatter_profiles_hernquist} shows the analytically calculated scatter rate for a constant 
($\sigma_T^{\rm max}/m_\chi=10~{\rm cm}^2\,{\rm g}^{-1}$, solid lines) and 
velocity-dependent ($v_{\rm max}=30~{\rm km}\,{\rm s}^{-1}$, $\sigma_T^{\rm max}/m_\chi=10~{\rm cm}^2\,{\rm g}^{-1}$, dashed lines) cross section. 
In the constant cross section case we neglect the velocity dependence, whereas the velocity-dependent case
assumes the parametrisation of Eq.~\ref{eq:cross}. We show these rates for $6$ different haloes, where we have chosen the scale radii
$a$ to follow a cosmologically motivated mass-concentration relation $c \propto M^{-1/9}$ \citep[e.g.][]{Neto2007}. In the other two panels
of Figure~\ref{fig:scatter_profiles_hernquist} we 
compare the analytic scatter rates to the result of the N-body ($N=10^6$, Plummer equivalent softening length $\epsilon=0.0063 \times a$) calculation assuming a frozen particle configuration. The middle panel 
shows the constant cross section, whereas the right panel shows the result for the velocity-dependent cross section. All cases
show good agreement between the analytic result and the simulation.
 
In Figure~\ref{fig:scatter_profile_hernquist_live} (two top panels) we show the density and velocity dispersion profiles for the same
haloes and the same constant cross section with a live N-body ($N=10^6$, $\epsilon=0.0063 \times a$) system after $25~t_0$, where
\begin{equation}
\rho_0 = \frac{M}{2 \pi a^3}, \quad v_0 = a \sqrt{4 \pi G \rho_0}, \quad t_0^{-1} = \frac{\sigma_T}{\sqrt{2}\pi} \hat{a} \sqrt{\frac{G M^3}{a^7}},
\end{equation}
and $\hat{a}=2.26$ \citep[see][]{KodaShapiro2011}. As expected all haloes evolve in the same way if time is expressed in units of $t_0$, density in units of $\rho_0$
and velocity in units of $v_0$. Core expansion is maximised around $25~t_0$, and after that the core-collapse phase slowly starts.
The two lower panels of Figure~\ref{fig:scatter_profile_hernquist_live} show density and velocity dispersion profiles for the velocity-dependent case after $1~{\rm Gyr}$. Within this time span the haloes reach different levels of core expansion due to their different velocity structure with the lowest mass
halo showing the largest core at that time.

\section{Cosmological simulations of a Galactic SIDM halo}\label{sims}

\begin{figure*}
\centering
\includegraphics[width=0.475\textwidth]{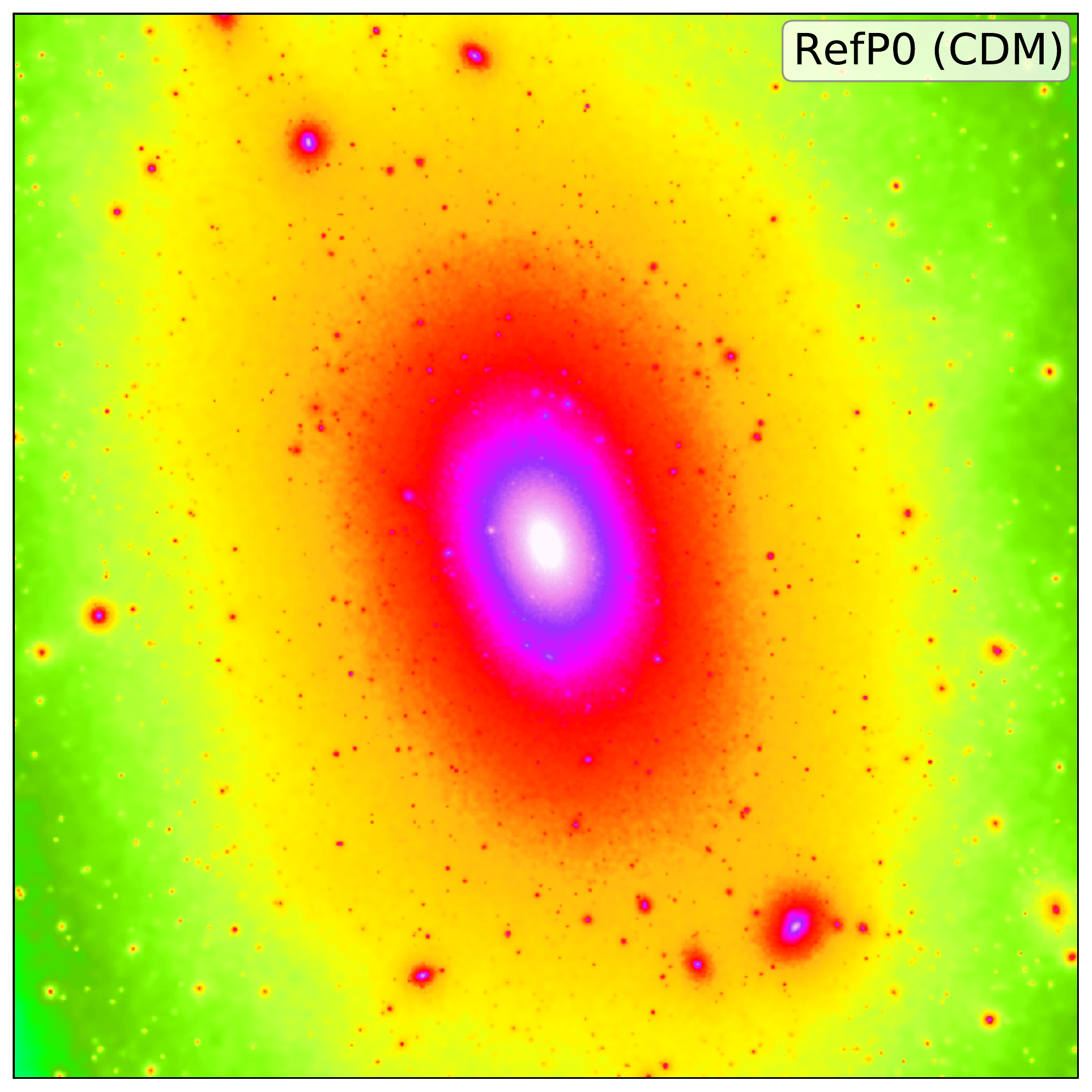}
\includegraphics[width=0.475\textwidth]{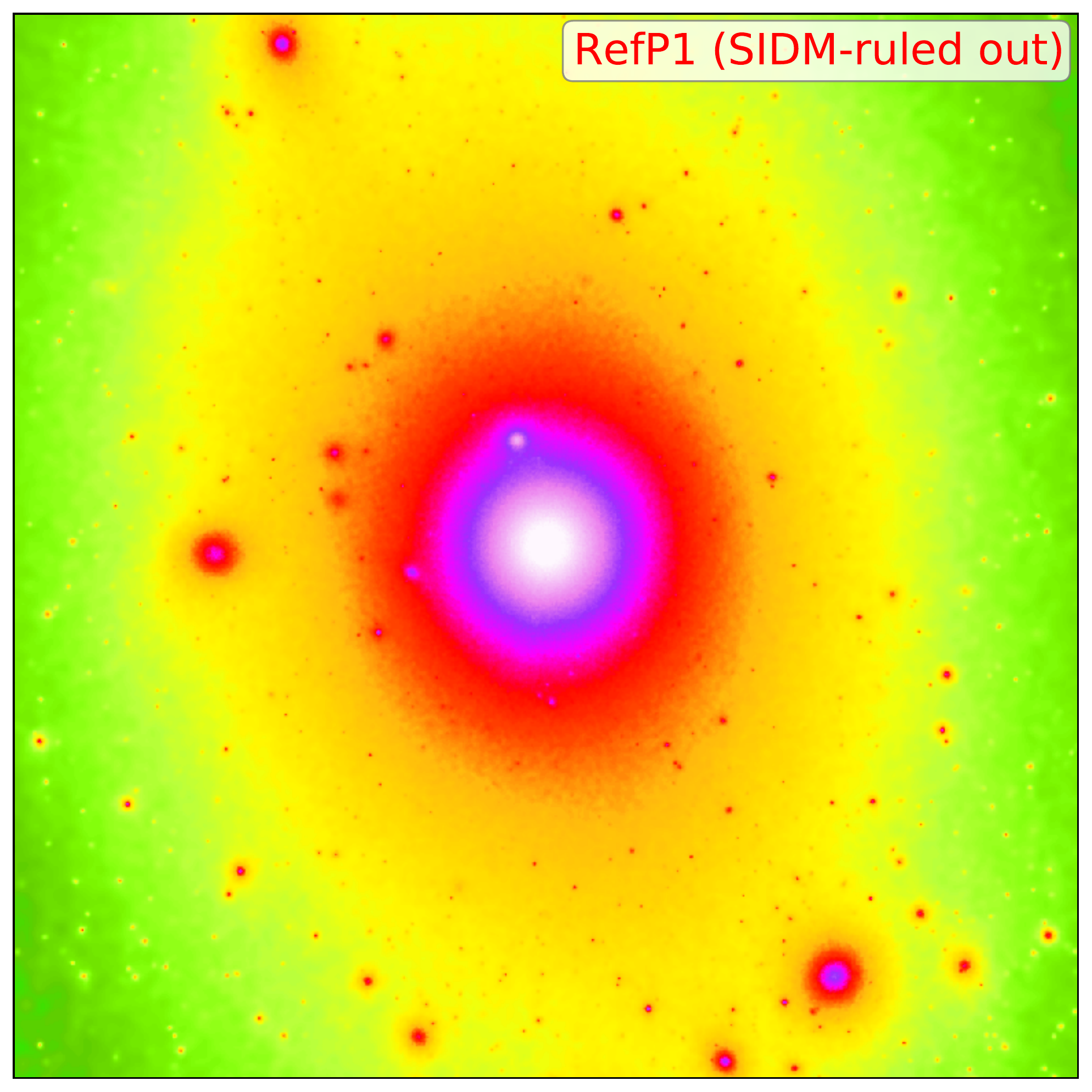}
\includegraphics[width=0.475\textwidth]{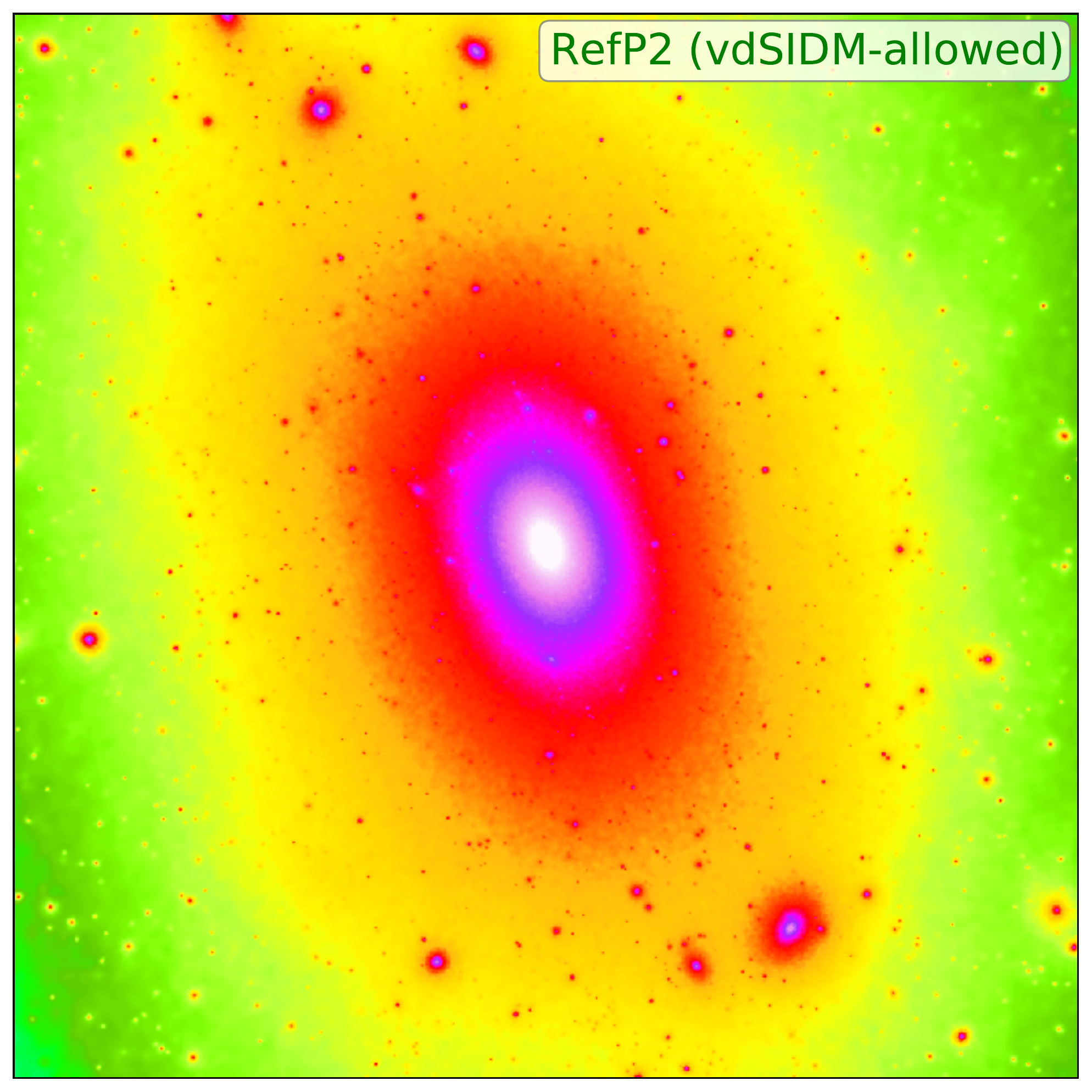}
\includegraphics[width=0.475\textwidth]{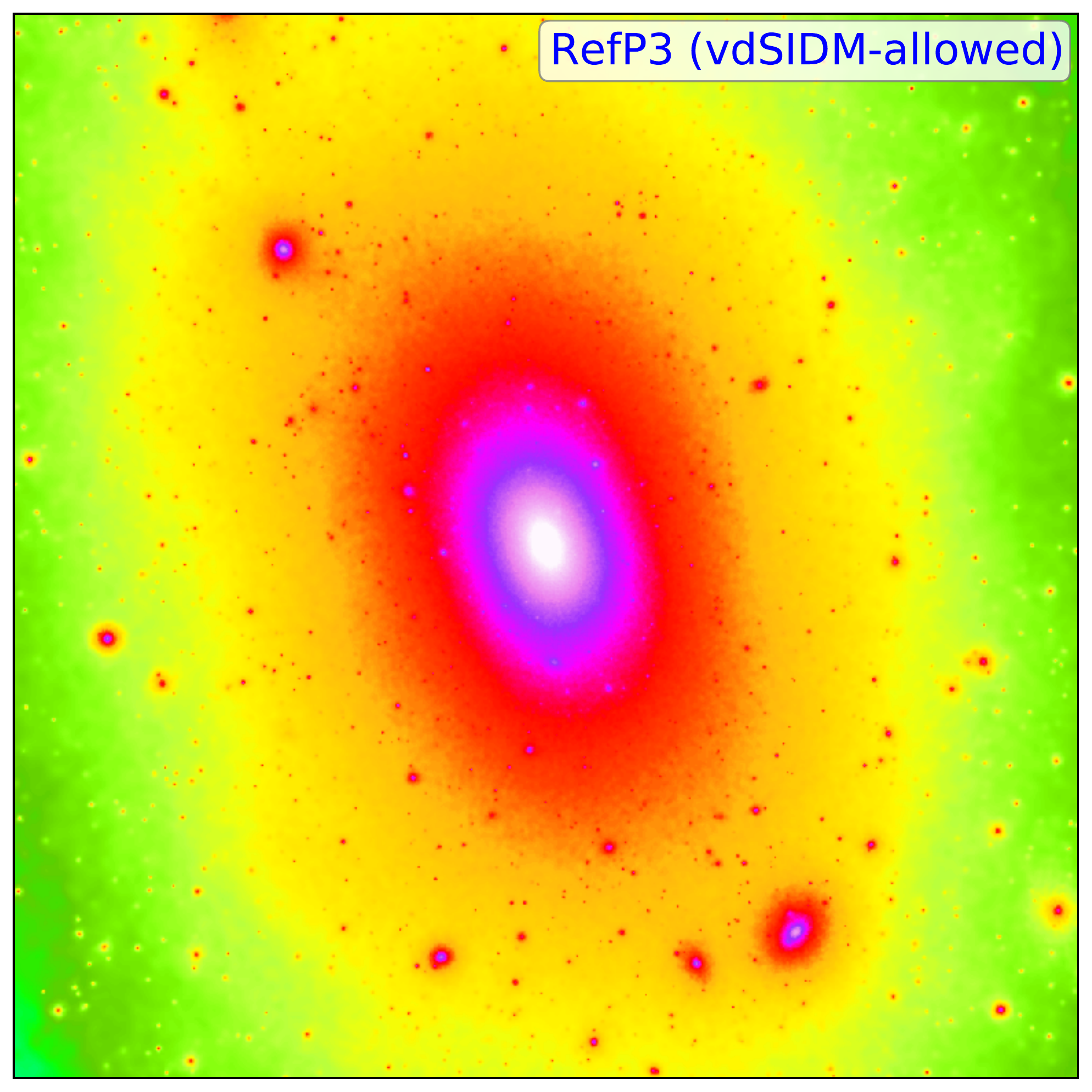}
\caption{Density projections of the Aq-A halo for the different DM models of Table~\ref{table:ref_points} (RefP0-3). The projection cube 
has a side length of $270$~kpc. Clearly,
the disfavoured RefP1 model with a large constant cross section produces a very different density distribution with a spherical core 
in the centre, contrary to the elliptical and cuspy CDM halo. Also, substructures are less dense and more spherical in this simulation. 
The vdSIDM models RefP2 and RefP3 on the other hand can hardly be distinguished from the CDM case (RefP0).}
\label{fig:projections}
\end{figure*}

\subsection{Initial Conditions and Models}\label{IC}

The initial conditions for our cosmological MW-like halo simulations are taken from the Aquarius Project \citep{Springel2008}. 
This has the advantage that we can compare subhalo properties object-by-object with CDM simulations that have 
already been carried out at very high resolution. In addition, we can directly compare our results with those of
\cite{Boylan2011b}, who 
found the massive subhalo failure problem described in the Introduction using Aquarius data\footnote{We note that \citet{Boylan2011b} also did the analysis
using data from the Via-Lactea simulations \citep{Diemand2008} and found the same problem.}. 
The Aquarius initial conditions use the following cosmological parameters: $\Omega_m=0.25$, $\Omega_{\Lambda}=0.75$, $h=0.73$,
$\sigma_8=0.9$ and $n_{s}=1$; where $\Omega_m$ and $\Omega_{\Lambda}$ are the
contribution from matter and cosmological constant to the mass/energy density
of the Universe, respectively, $h$ is the dimensionless Hubble constant
parameter at redshift zero, $n_s$ is the spectral index of the primordial
power spectrum, and $\sigma_8$ is the rms amplitude of linear mass fluctuations
in $8~h^{-1}\,{\rm Mpc}$ spheres at redshift zero. 
The haloes were selected to be representative of the MW halo with different merger histories. 
Here we focus on the Aq-A halo and re-simulate it within the SIDM model described before at different resolutions.
We note that we only consider scatter events between high resolution particles, i.e. massive boundary particles
behave like CDM. This does not introduce any bias effects since our SIDM models produce the same large scale
structures as CDM, i.e. tidal effects due to the boundary particles are treated correctly even when neglecting
their self-scattering. Furthermore the high resolution regions of all the simulations presented here are free of
any contamination from boundary particles. 

In Table~\ref{table:ref_points} we show the SIDM models we consider in this paper. RefP0 is the vanilla CDM case and
RefP1 only serves as a test case to demonstrate various effects since is ruled out based on astrophysical constraints.
For instance, observations of the Bullet cluster place an upper limit to the DM scattering cross section which is below the
value used in the RefP1 case: $\sigma_T/m_{\chi}<1.25~{\rm cm}^2~{\rm g}^{-1}$ \citep{Randall2008}\footnote{A large constant 
cross section model will also create a core
and isotropise the central part of galaxy clusters. There are stringent constraints on the cross section for these systems,
for instance, \citet{Miralda2002} derived a limit of $\sigma_T/m_{\chi}<0.02~{\rm cm}^2~{\rm g}^{-1}$ by analysing the ellipticity of a cluster core.}. RefP2 and RefP3 on the other hand,
do not violate any constraints and potentially have a significant effect on the density profiles of low-mass subhaloes. The latter two 
reference points are therefore the ones we will mainly focus on. 

\begin{table}
\begin{tabular}{cccc}
\hline
Name         & Type               & $\sigma_T^{\rm max}/m_\chi$ $[{\rm cm}^2\,{\rm g}^{-1}]\!\!\!\!\!$     & $v_{\rm max}$ $[{\rm km}\,{\rm s}^{-1}]\!\!\!\!$     \\
\hline
\hline
RefP0        & CDM                & /                                                        & /                                           \\  \hline 
RefP1        & SIDM (ruled out)   & $10$                                                     & /                                           \\  \hline 
RefP2        & vdSIDM (allowed)   & $3.5$                                                    & $30$                                        \\  \hline 
RefP3        & vdSIDM (allowed)   & $35$                                                     & $10$                                        \\ 
\hline
\end{tabular}
\caption{Reference points and their particle physics parameters explored in our simulations. RefP1 serves only as
a benchmark point for tests, since it is well-known that such a large constant cross section violates various
astrophysical constraints. RefP2 and RefP3 do not violate any constraints and potentially have a significant effect 
on the density profiles of low-mass subhaloes. The latter two reference points are therefore the ones we will mainly focus on.}
\label{table:ref_points}
\end{table}

In the following, we investigate the effects of DM self-interaction on the present-day properties of the MW-like halo and its 
subhaloes, which consist of self-bound groups of particles identified by the SUBFIND algorithm 
\citep{Springel2001} down to a group with a minimum of 20 particles.

\subsection{Main halo}\label{main_halo}

\begin{figure*}
\centering
\includegraphics[width=0.475\textwidth]{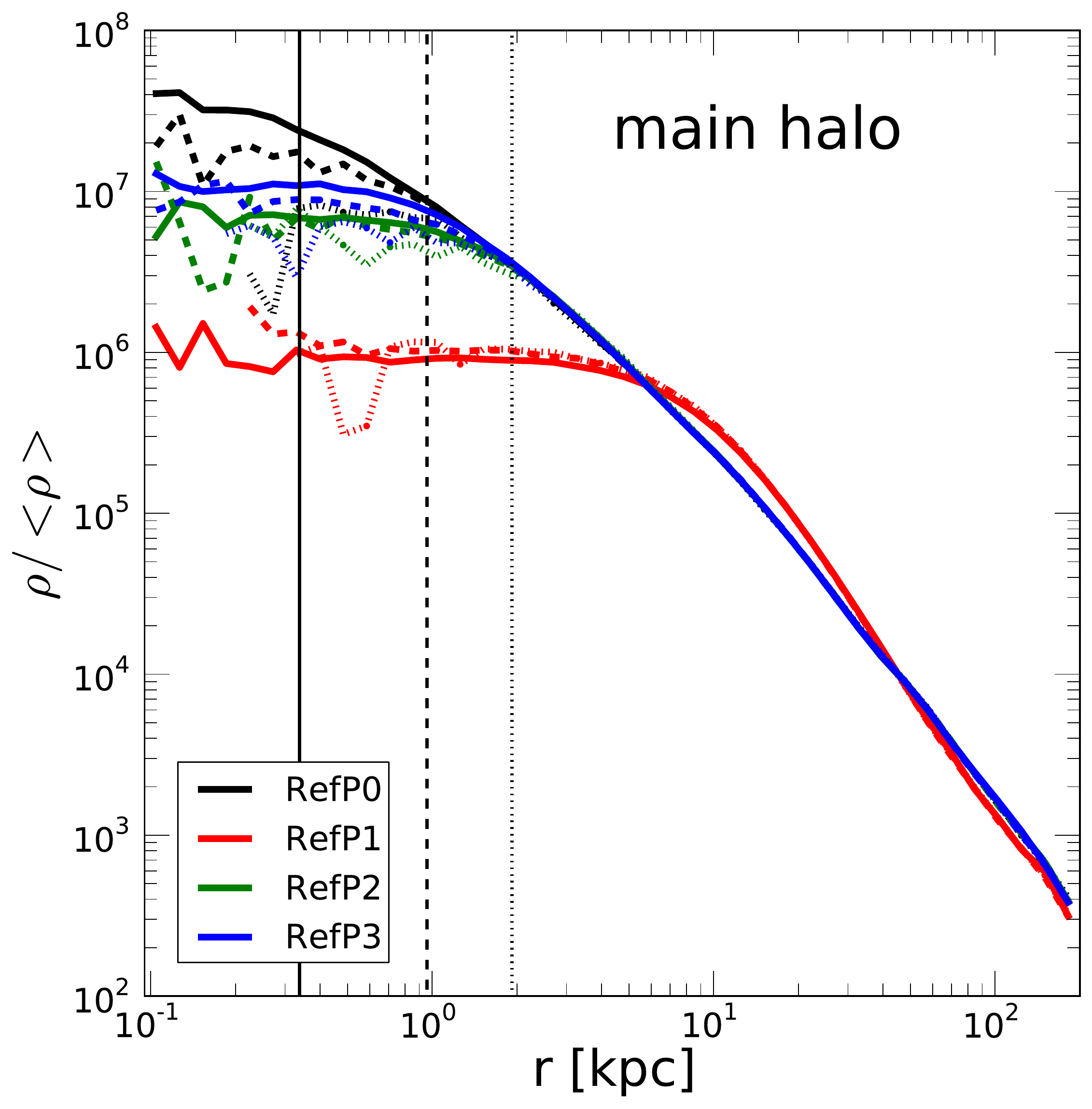}
\includegraphics[width=0.475\textwidth]{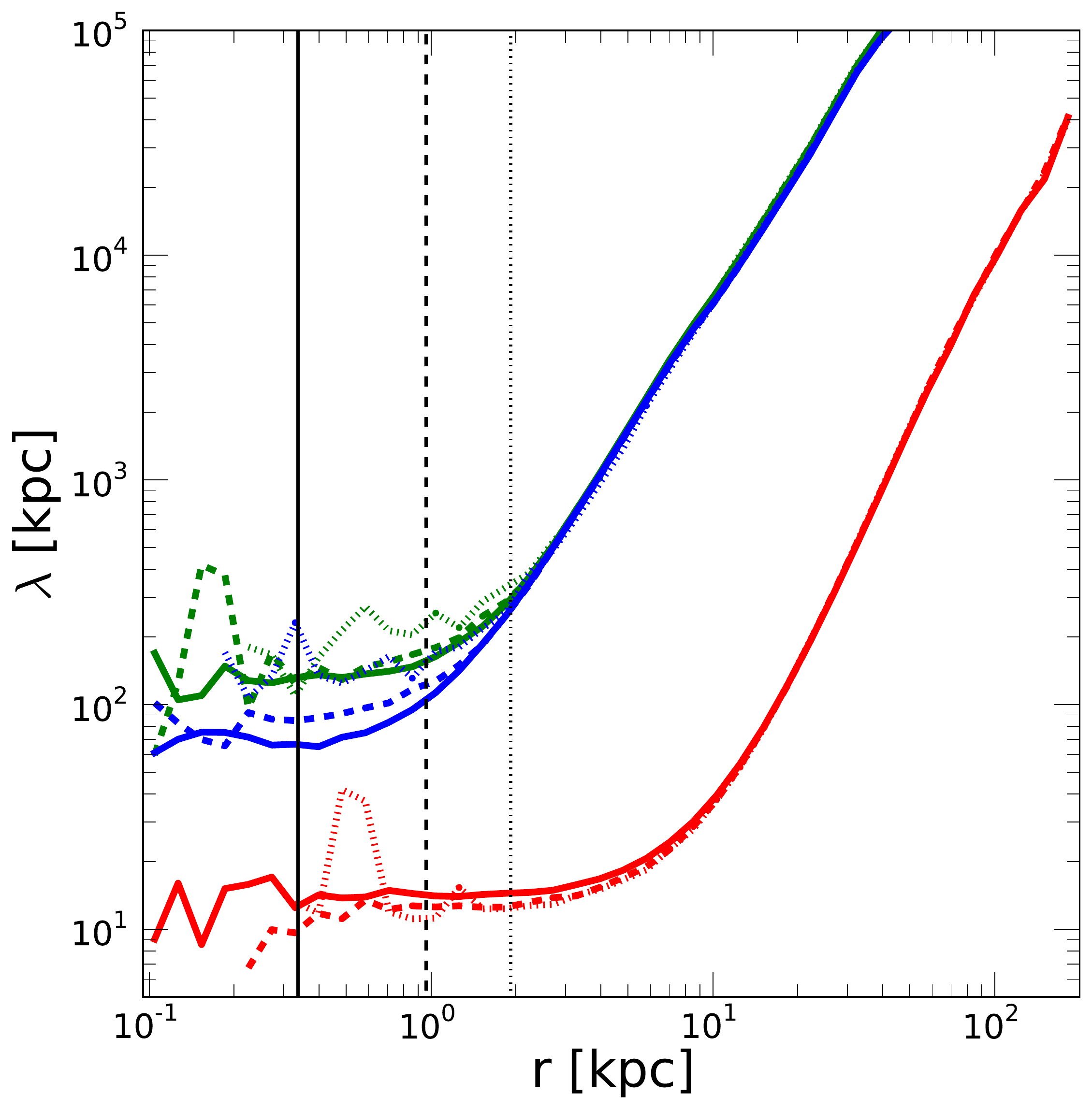}
\caption{Left panel: Density profile of the Aquarius Aq-A main halo at resolution levels 5 (dotted), 4 (dashed) and 3 (solid) for 
the different SIDM reference points we consider (see Table~\ref{table:ref_points}) as shown in the legend. Right panel: Mean free path
as a function of radius for the SIDM models we used. The softening
length ($2.8$ times Plummer equivalent) of each resolution is marked by a vertical black line. The standard CDM profiles (RefP0) 
are shown in black. We achieve good convergence in all our runs, with the inner profiles changing significantly depending on the DM 
model employed. Clearly, RefP1 produces the largest difference having a large core due to the constant scattering cross section. We note that 
this model is ruled out by current astrophysical constraints and is shown here just as reference.}
\label{fig:mainhalo_density}
\end{figure*}

A first visual impression of the structure of the halo in the different DM models is given in Figure~\ref{fig:projections}, where we show
density projections of the halo for the models given in Table~\ref{table:ref_points}. Clearly, the disfavoured RefP1 model with a large constant 
cross section has a very different density distribution with a spherical core in the centre. Substructures also share
these properties being less dense and more spherical in this simulation. The vdSIDM models RefP2 and RefP3 on the other hand can 
hardly be distinguished from the CDM case (RefP0) in this figure. As we see below, the internal structure of the massive subhaloes does
actually change significantly with respect to the CDM case. We note that the spherical cores in self-interacting models are due to the assumed isotropic
scattering process, which tends to isotropise particle orbits leading to a more isotropic and spherical configuration.

The left panel of Figure~\ref{fig:mainhalo_density} shows the main halo density profiles for the different models. whereas the right panel shows the mean free path 
$\lambda=(\rho\left<\sigma_T/m_{\chi}\right>)^{-1}$
as a function of radius for the SIDM models. The dotted, dashed and solid lines show different levels of resolution, characterised
by a particle mass $m_p$ and a Plummer equivalent gravitational softening length $\epsilon$: 
Aq-A-5 ($m_p=3.143\times10^6$~M$_{\odot}$, $\epsilon=684.9$~pc), Aq-A-4 ($m_p=3.929\times10^5$~M$_{\odot}$, $\epsilon=342.5$~pc) and 
Aq-A-3 ($m_p=4.911\times10^4$~M$_{\odot}$, $\epsilon=120.5$~pc). The runs show good convergence for radii larger than $2.8\epsilon$ indicated 
by the vertical lines. 

\begin{figure}
\centering
\includegraphics[width=0.475\textwidth]{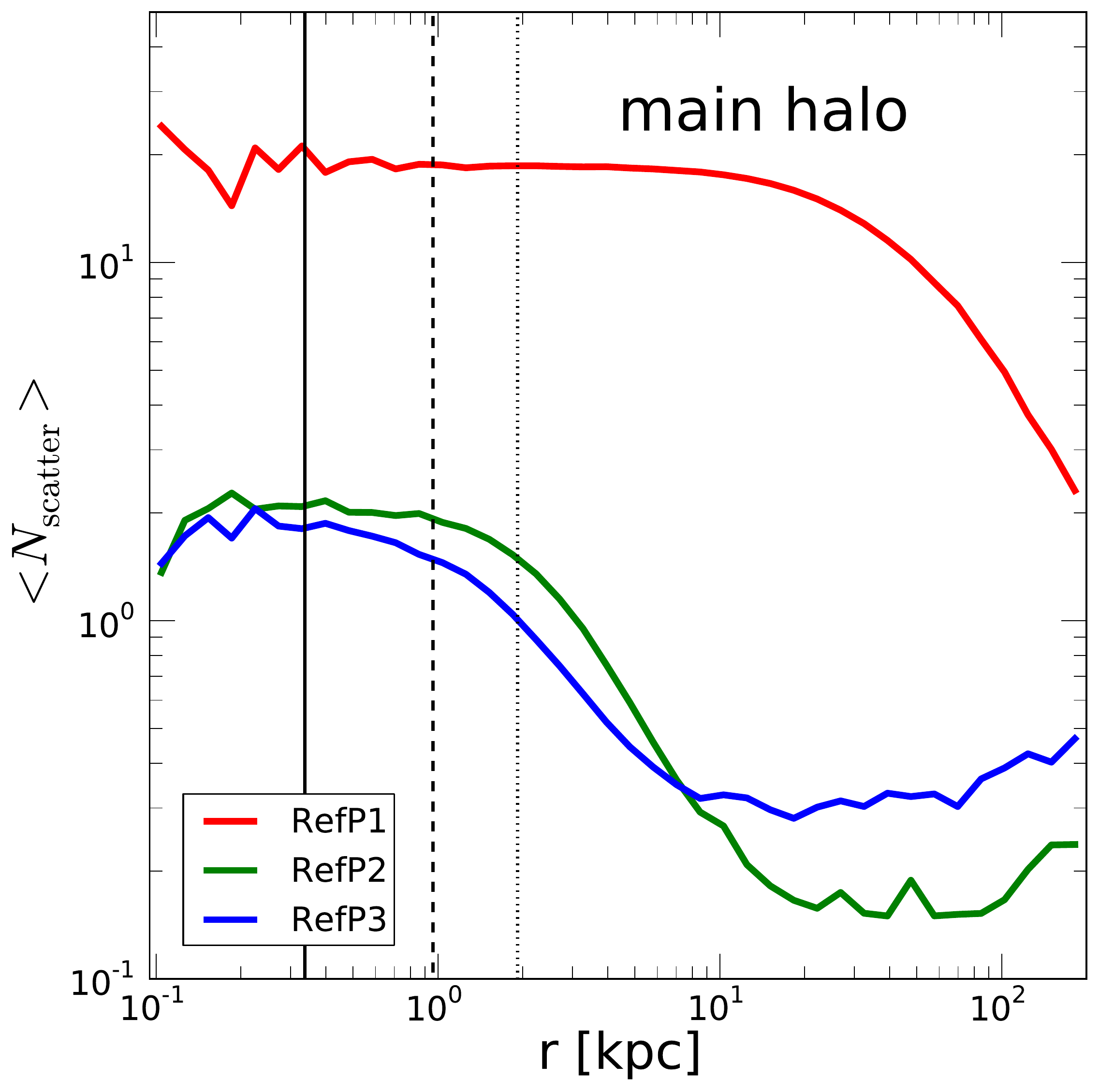}
\caption{Radial profiles of the mean number of scatter events for the Aq-A-3 simulations RefP1, RefP2 and RefP3. The large and constant cross-section 
of RefP1 produces a significantly higher number of scatter events at a given radius compared to RefP2 and RefP3. We note that the vdSIDM points
RefP2 and RefP3 lead to slightly different scatter profiles.
}
\label{fig:mainhalo_scatter}
\end{figure}

In the figure we see that RefP1 develops a large core reaching the solar circle ($\sim7$~kpc). This is because the cross section
has no velocity dependence in this case and the particle scattering works at full strength irrespective of (sub)halo mass. Although
this case is ruled out by current astrophysical constraints (see Section \ref{mod_sec}), it serves as a reference for the effect of a 
large scattering cross section at the scales of MW-like haloes in a full cosmological simulation. On the contrary, RefP2 and RefP3 result
in a main halo whose density profile follows very closely the one from the CDM prediction of RefP0 down to $1$~kpc from the centre.
At smaller radii, where the typical particle velocities are smaller, self-interaction is large enough to produce a core. The mean free path
radial profile clearly illustrates the radius where collisions are more important for the different SIDM models, which is around the core
radius. It also highlights the difference between the RefP2 and RefP3 models, with the former having a larger core than the latter, because
its self-interaction cross section peaks at a larger velocity dispersion (occurring at larger radii) despite of having a lower value of 
$\sigma_T/m_{\chi}$.

\begin{figure*}
\centering
\includegraphics[width=0.475\textwidth]{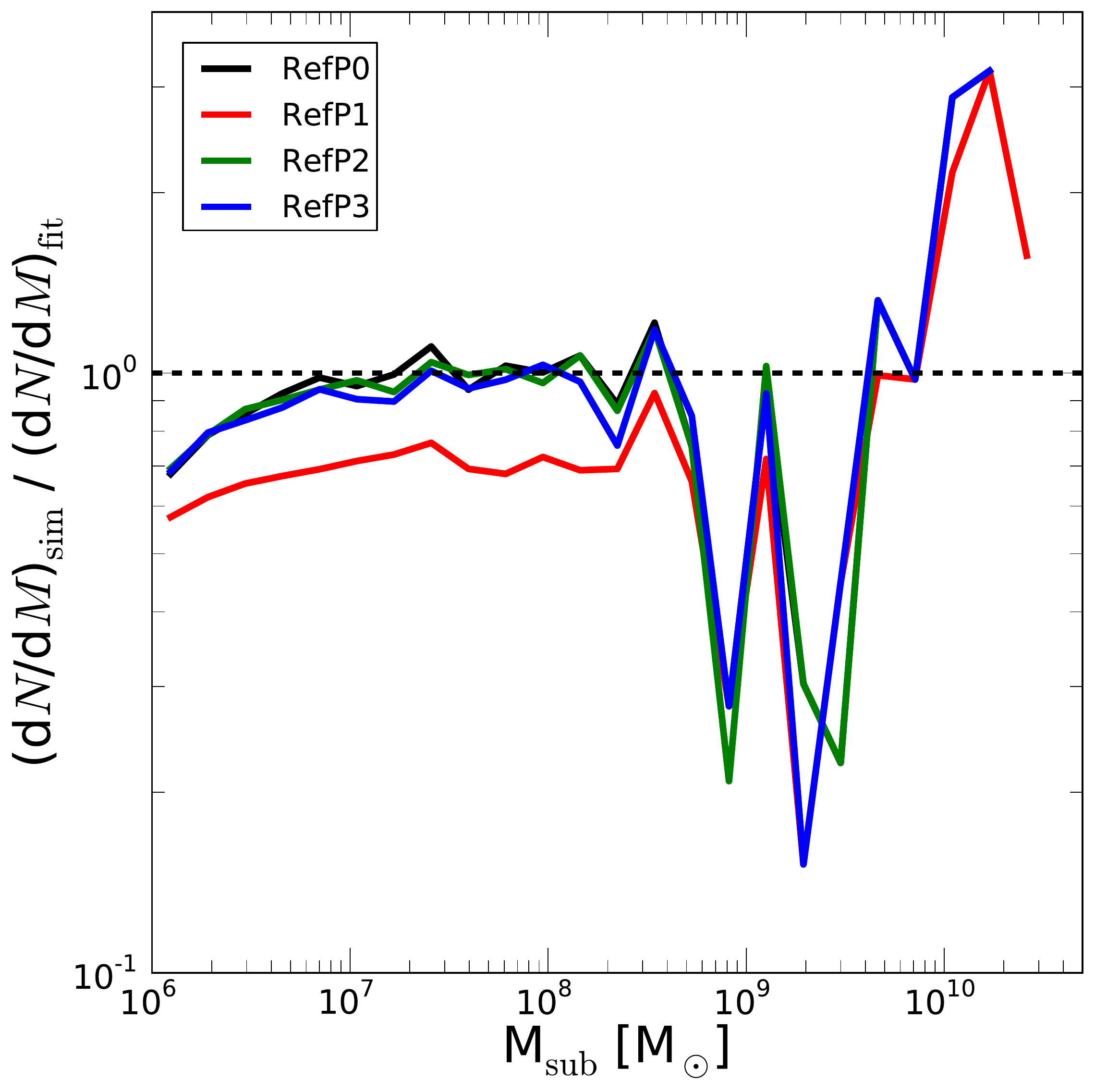}
\includegraphics[width=0.495\textwidth]{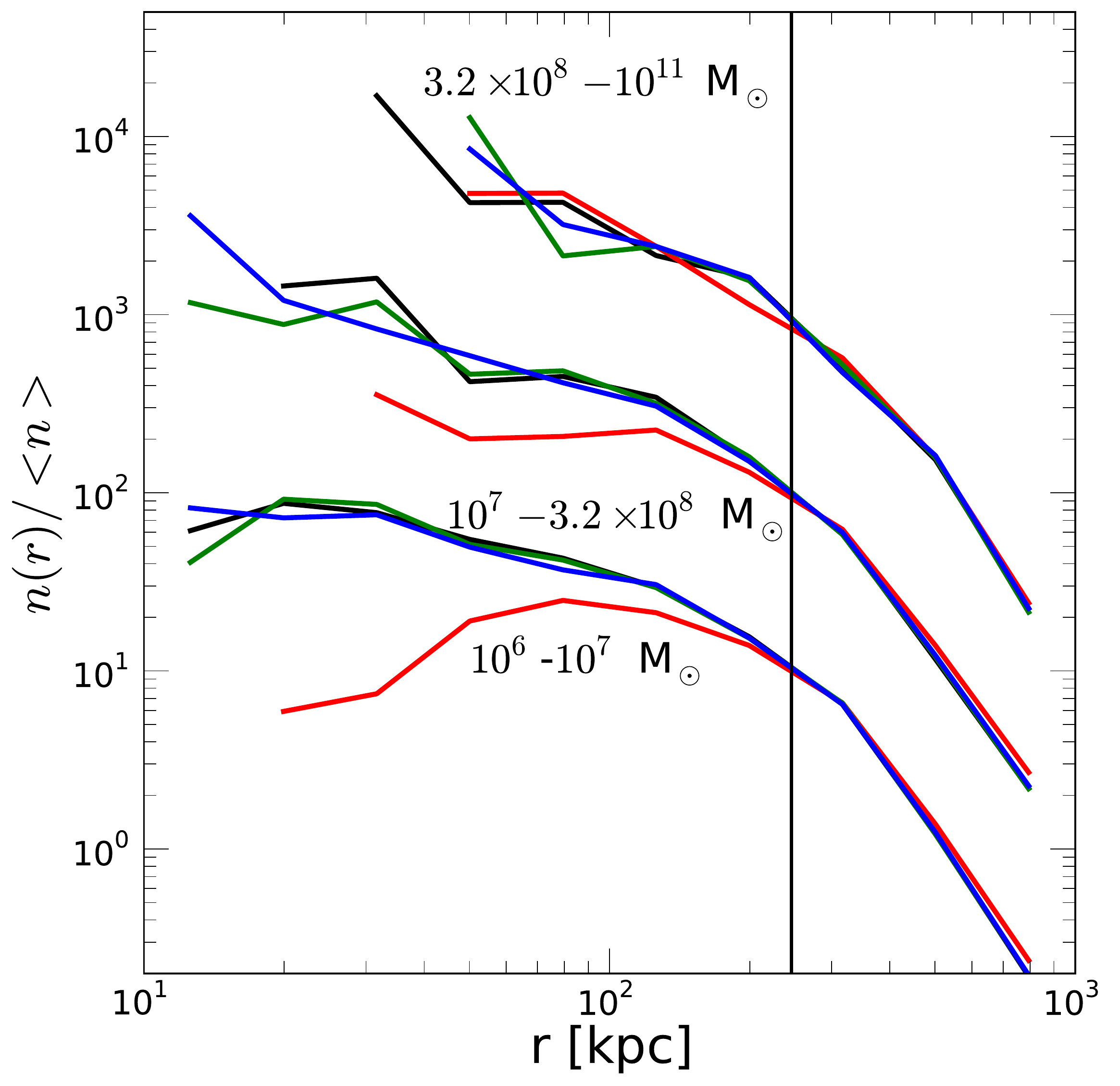}
\caption{Left panel: Ratio of the subhalo mass function for our different SIDM reference models to the best fit subhalo mass function of \citet{Springel2008}.
Right panel: radial distribution of the subhalo number density (the virial radius is marked with a solid black line). The
subhalo population has been divided in three mass bins: $(10^6,10^{7})~$M$_{\odot}$, $(10^7,3.2\times10^{8})~$M$_{\odot}$ and $(3.2\times10^{8},10^{11})~$M$_{\odot}$, from bottom to top, respectively (the last two bins are shifted up by one (two) dex for clarity).
The constant cross section SIDM model
(RefP1) gives a significantly different subhalo abundance, particularly in the inner regions of the main halo, than the CDM model (RefP0). On the contrary,
the velocity-dependent models (RefP2 and RefP3) are nearly indistinguishable from the collisionless case. For clarity, we only include the results
for the highest resolution simulations (level 3).}
\label{fig:sub_stat}
\end{figure*}

Each simulation particle records the total number of scatter events during its dynamical evolution. We can use this information to show 
alternatively the size of the collisional radius by constructing a radial
profile of the mean number of scatter events as a function of radius. This is shown in Figure~\ref{fig:mainhalo_scatter} for the highest
resolution Aq-A-3 simulations for RefP1, RefP2 and RefP3. Clearly, the large and constant cross-section of RefP1 produces a significantly 
higher number of scatter events at a given radius compared to RefP2 and RefP3. 

\subsection{Subhaloes}\label{subhaloes}

Our main focus in this work is the structural change of the subhalo population in a SIDM halo. In the following we will mainly focus on
the subhalo population within $300$~kpc halocentric distance. 
In the left panel of Figure~\ref{fig:sub_stat} we first show the ratio of the subhalo mass function of our different models (for the highest 
resolution run, level 3) to the best fit subhalo mass function of the CDM Aquarius haloes \citep[see][]{Springel2008}. 
The disfavoured RefP1 model 
differs significantly from the CDM prediction, contrary to the RefP2 and RefP3 models that have mass functions very 
similar to the RefP0
model. This means that subhalo masses do not change significantly in the vdSIDM models, whereas subhaloes lose mass in
the RefP1 model. The right panel of Figure~\ref{fig:sub_stat} shows the subhalo radial number density profiles for the different models. The
subhalo population has been divided in three mass bins: $(10^6,10^{7})~$M$_{\odot}$, $(10^7,3.2\times10^{8})~$M$_{\odot}$ and $(3.2\times10^{8},10^{11})~$M$_{\odot}$, 
from bottom to top, respectively (the last two bins are shifted up by one (two) dex for clarity, in fact
the three curves lie on top of each other) . Once again, the velocity-dependent scenarios are essentially indistinguishable from 
the CDM model (at all masses), whereas the RefP1 model shows a considerably lower abundance of subhaloes within the virial radius of 
the main halo (particularly for the least massive subhaloes), which is more acute in the inner regions. 
Although not shown here, we checked that the lower resolution runs converge to the level 3 curves for both subhalo mass function
and radial distribution for all models.

\begin{figure*}
\centering
\includegraphics[width=0.475\textwidth]{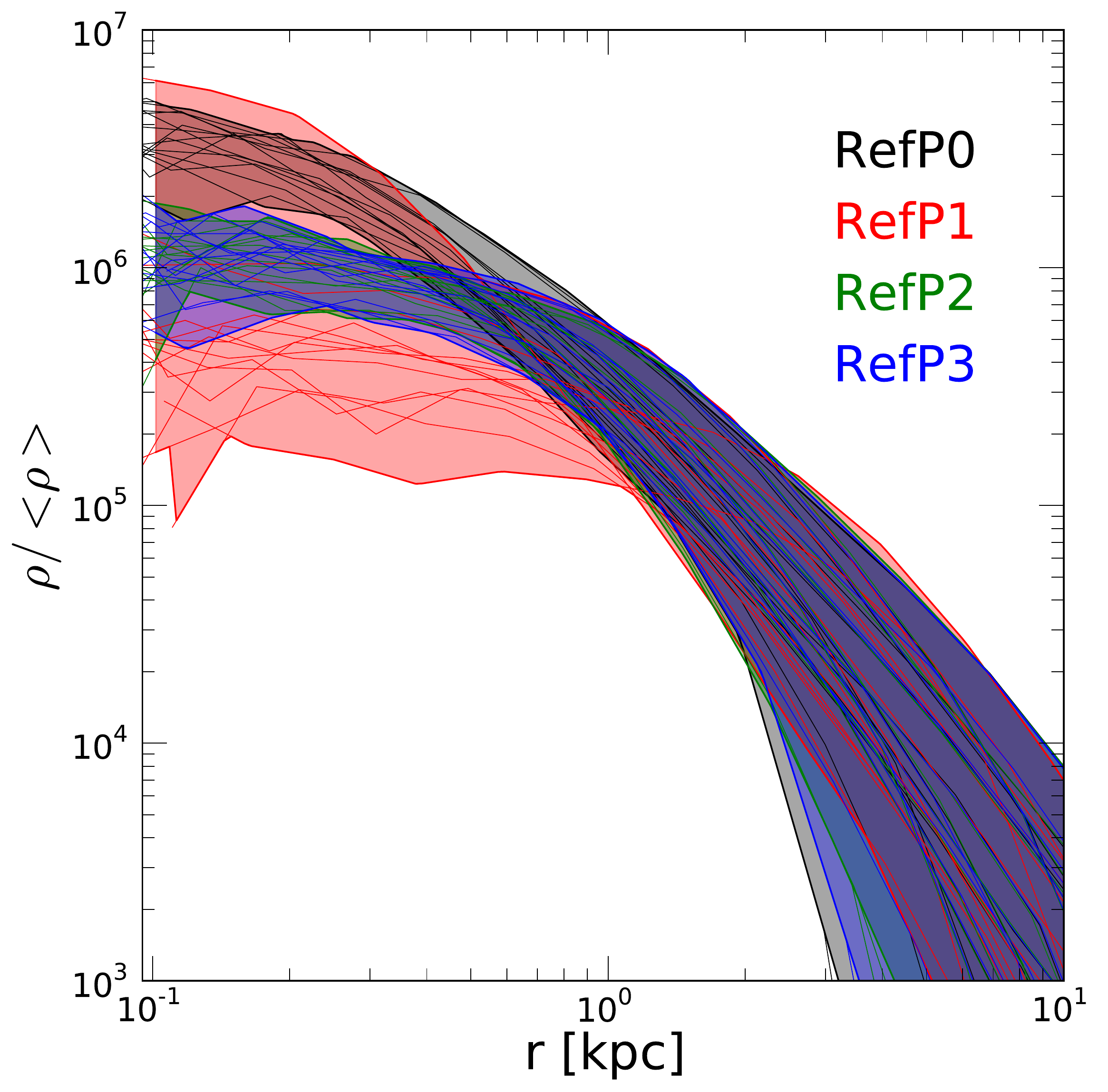}
\includegraphics[width=0.475\textwidth]{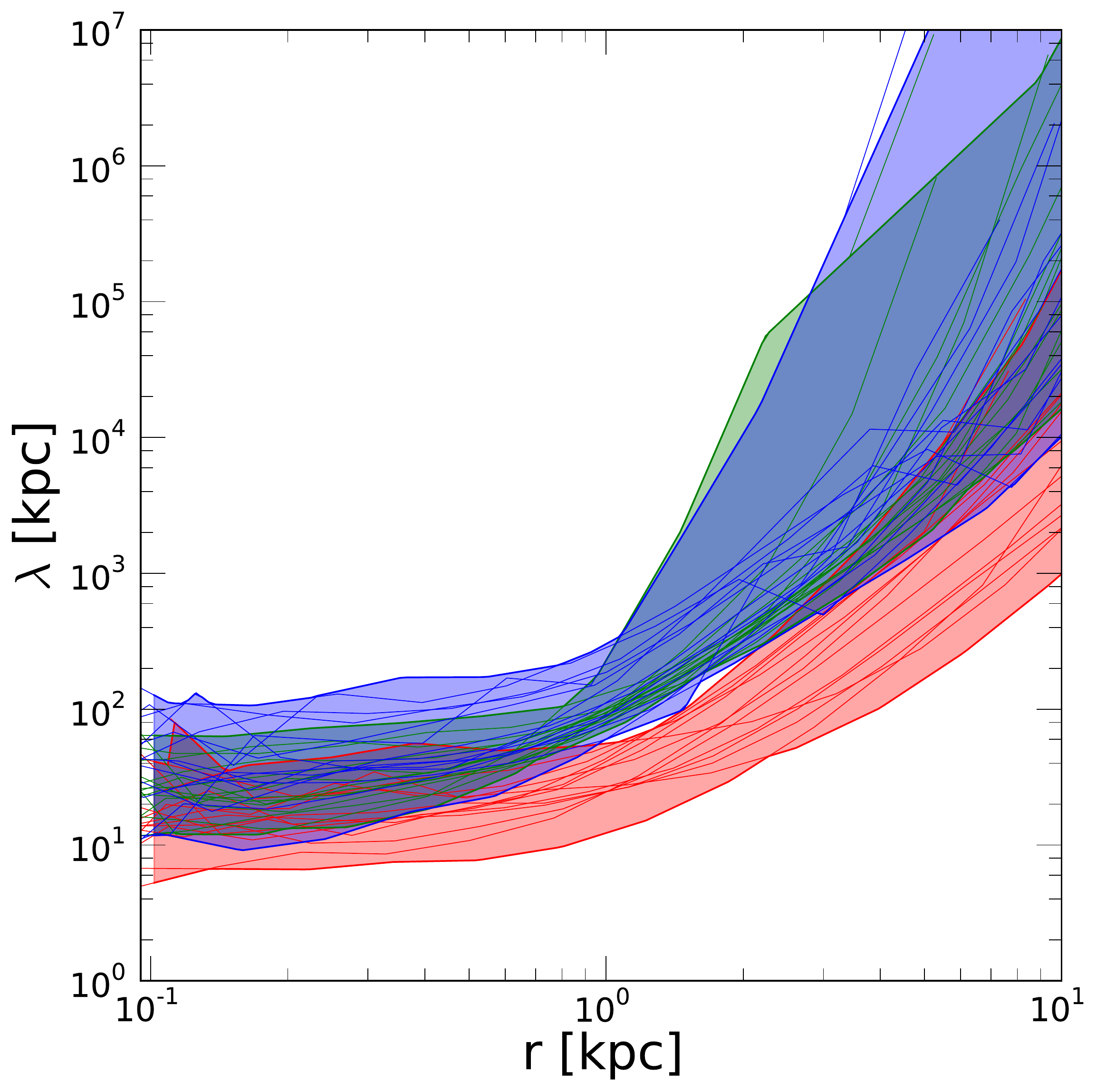}
\caption{Subhalo density (left panel) and mean free path (right panel) at $z=0$ for the top 15 most massive subhaloes (largest peak circular velocity) in our highest resolution
simulations (level 3). The vdSIDM models produce subhaloes with cores of approximately $600$~pc and there is little difference between RefP2 and RefP3, with the
latter having slightly larger cores. Beyond $1$~kpc, these cases are indistinguishable from CDM profiles. The case where self-interaction is
independent of velocity (RefP1) results in cores that have lower density and are more extended (approximately a factor of $\sim$2). We note that one of the shown subhaloes of RefP1 entered already the core-collapse regime resulting in a very steep density profile, whereas most of the other subhaloes are all still in the core-expansion phase.}
\label{fig:sub_density}
\end{figure*}

The subhalo evaporation seen in the RefP1 model is explained by the interaction that takes place between the
particles of a given subhalo and the higher velocity particles of the surrounding host halo. Since these collisions are elastic, the net effect
is a transfer of energy to the particles in the subhalo, unbinding the ones with the lowest binding energies. These interactions are 
more common in the inner region of the main halo where the density is higher, and strongly reduce the abundance of subhaloes there. On the
other hand, evaporation is not effective in the velocity-dependent models where self-interactions are strongly suppressed for high relative 
velocities, typically expected between particles in the subhaloes and those of the host. This is an important result since it implies
that the overabundance of dwarf haloes is still present for the vdSIDM models 
explored here. This means that the 
``missing satellite problem'' would need to be solved by invoking astrophysical processes as is done for the CDM model. We note however that this
is only true for elastic scattering, for the inelastic case, subhalo evaporation should have a relevant role. We leave the study of this case
for a future analysis.

\begin{figure*}
\centering
\includegraphics[width=0.475\textwidth]{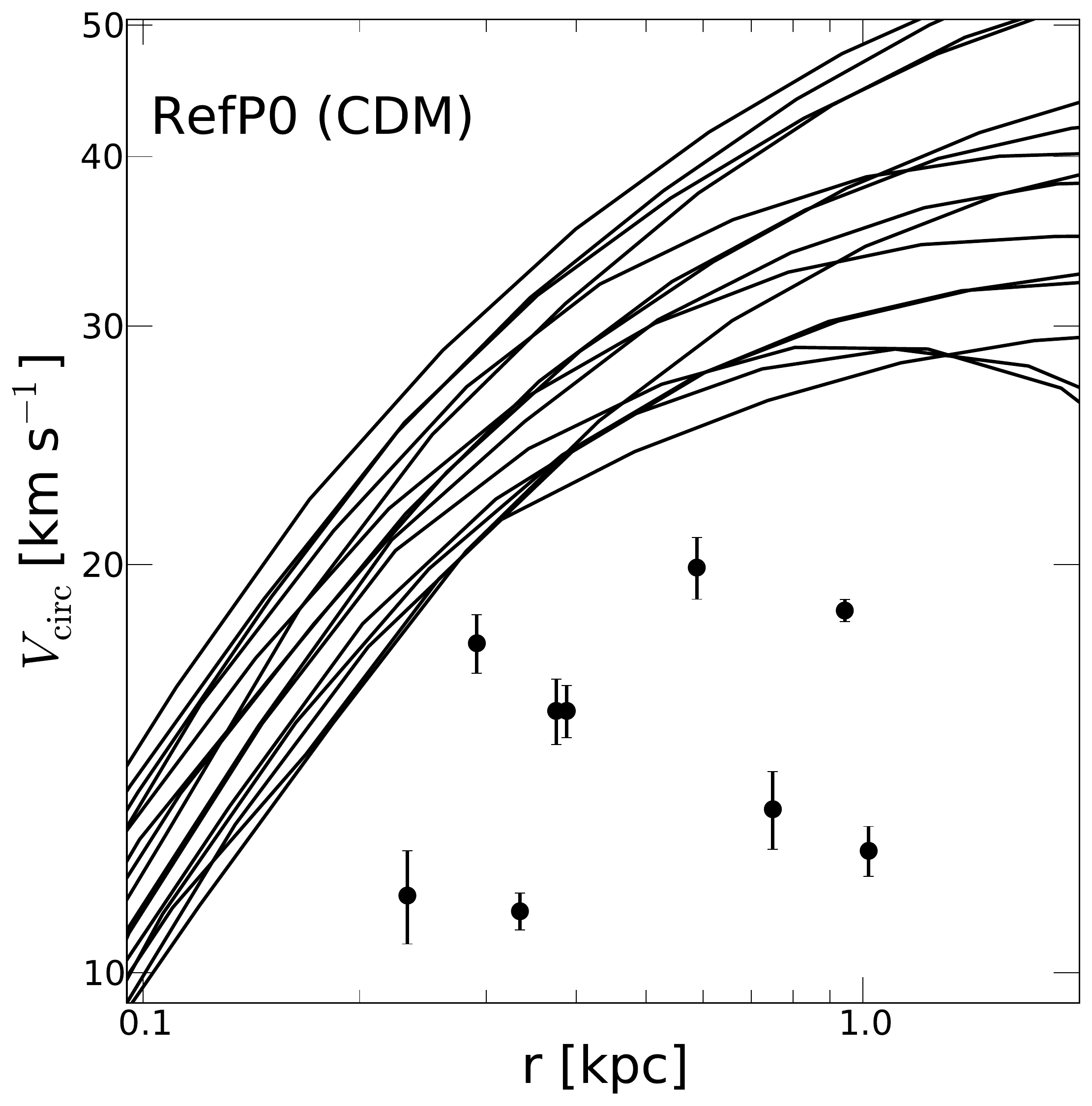}
\includegraphics[width=0.475\textwidth]{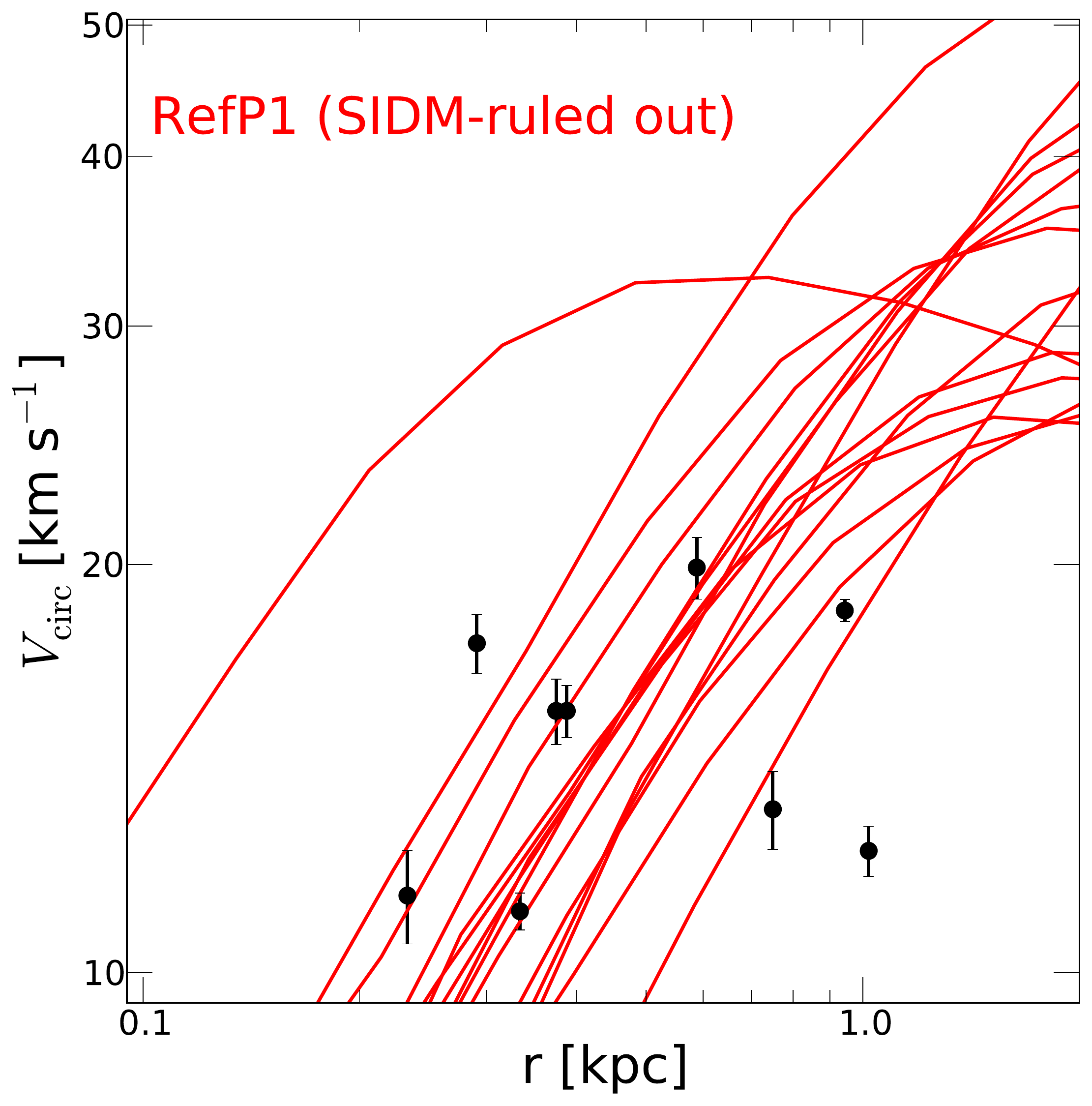}
\includegraphics[width=0.475\textwidth]{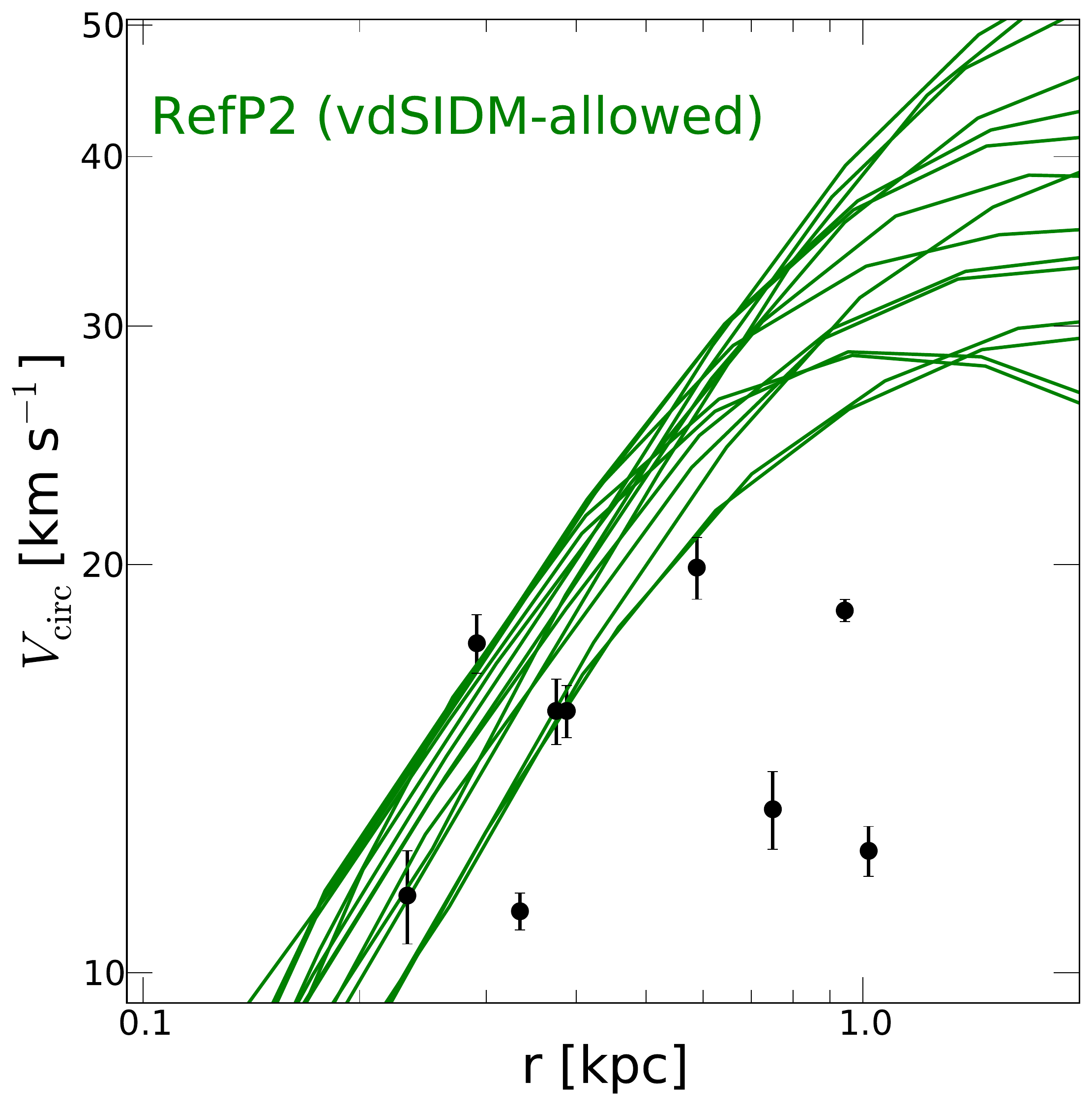}
\includegraphics[width=0.475\textwidth]{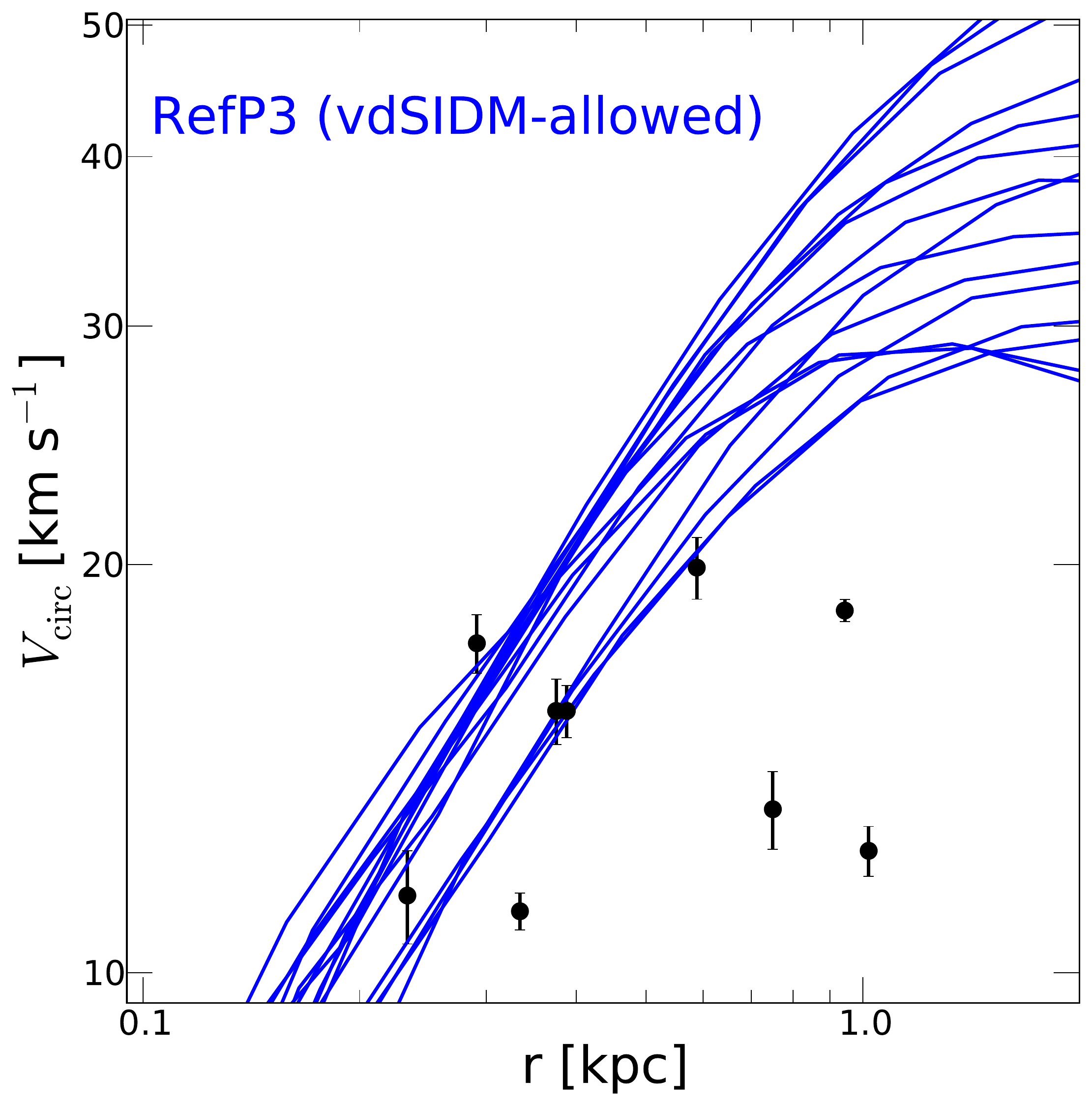}
\caption{Circular velocity profiles at $z=0$ for the top 15 most massive subhaloes (largest peak circular velocity) of the Aquarius-A halo
for the different SIDM reference models as
given in the legends. The upper left panel shows the standard CDM case, while the bottom panels show two examples of the
vdSIDM models described in section \ref{mod_sec}. Observational estimates of $V_{\rm circ}(r_{1/2})$ for the MW dSphs are
shown with black circles with error bars \citep{Walker2009, Wolf2010}. All SIDM results are shown at level 3 resolution which is sufficient
for convergence due to the subhalo density cores that form in these models
(see Figures~\ref{fig:sub_density} and \ref{fig:sub_conv}). RefP0 is shown at level 2 resolution ($2.8 \times 65.8 \sim 184$~pc spatial resolution),
because the CDM subhaloes form cuspy profiles which require higher numerical resolution for convergence (see Figure~\ref{fig:sub_conv}).
Clearly, the most massive subhaloes in the CDM model are dynamically inconsistent
with the MW dSphs, whereas the SIDM subhaloes are consistent with the data. We note that the constant cross section RefP1 case is ruled out by different
observations at the scale of galaxy clusters and is shown here only as a reference. One of the shown subhaloes of RefP1 entered already the core-collapse regime clearly visible from the circular velocity profiles (see also Figure~\ref{fig:sub_density} for the corresponding steep density profiles).}
\label{fig:sub_vcirc}
\end{figure*}

The internal structure of the subhalo population can be appreciated in the left panel of Figure~\ref{fig:sub_density} that shows the density profiles 
at $z=0$ for the 15 
most massive subhaloes
in our highest resolution simulations (level 3) for the cases RefP0 (black), 
RefP1 (red), RefP2 (blue) and RefP3 (green).  The figure shows that the 
vdSIDM cases produce subhaloes with cores of approximately $600$~pc; there is little difference between cases RefP2 and RefP3, with the
latter having slightly larger cores. Beyond $1$~kpc, these cases are indistinguishable from CDM. The case where self-interaction is 
independent of velocity (RefP1) results in cores that have lower density and are more extended (approximately a factor of $\sim$2). The
right panel of Figure~\ref{fig:sub_density} shows the mean free path as a function of radius for the difference SIDM cases. We note that
one of the shown subhalo of the RefP1 halo entered already the core-collapse regime, and therefore shows a very steep density profile. Other subhaloes
are still in the core-expansion phase resulting in significantly shallower profiles compared to the CDM (RefP0) curves.

We note that the cores of the main halo and the massive subhaloes (left panel of Figure~\ref{fig:mainhalo_density} and Figure~\ref{fig:sub_density}) are roughly of the same size.
Contrary to previous velocity-dependent SIDM models, ours seem to predict no strong scaling of the core size with mass. A concern might be that for 
cluster-sized systems, the hierarchical nature of structure formation implies that they were assembled from small mass halos that formed at earlier epochs. This means 
it is possible that although our vdSIDM models predict a vanishing cross-section for cluster scales, such systems might still exhibit large cores due to the
assembly of smaller significantly cored systems; such a large cluster core might violate observational constraints. However,the mass contribution
from these smaller systems to the central region of the cluster is expected to be subdominant \citep[][]{Gao2006} and therefore the evolution of
the inner regions of larger haloes is likely to be affected only in a minor way by the merging of smaller cored subhaloes. 

\section{Subhalo population: comparison with the bright MW dwarf spheroidals}\label{dsphs}

To check the the consistency between the subhalo population of our SIDM simulations and the kinematic data of the MW dSphs we construct
circular velocity curves for the most massive subhaloes within $300$~kpc halocentric distance for RefP0-3.
The dSphs sample consists of the 9 galaxies used in \citet{Boylan2011b}:
Fornax, Leo I, Sculptor, Leo II, Sextans, Carina, Ursa Minor, Canes Venatici I and Draco, selected with the criterion
$L_V>10^5$M$_{\odot}$. The Sagittarius dwarf was removed from this sample since it is in the process of interacting strongly 
with the galactic disc. This sample of bright dSphs (plus Sagittarius) is complete within the virial radius of the MW (excluding the possibility
of undiscovered systems hidden in the galactic plane). The Large and Small Magellanic Clouds are considerably brighter than the dSphs, 
and also more massive; \citet{Boylan2011b} puts a conservative lower limit to the maximum rotational velocity of the subhaloes that
could host these systems: $V_{\rm max}>40$~kms$^{-1}$. Although in this work we do not attempt to find Magellanic Cloud analogues in 
our simulations, we remark that some of the most massive subhaloes would actually correspond to these systems rather than
to the MW dSphs.

\begin{figure*}
\centering
\includegraphics[width=0.475\textwidth]{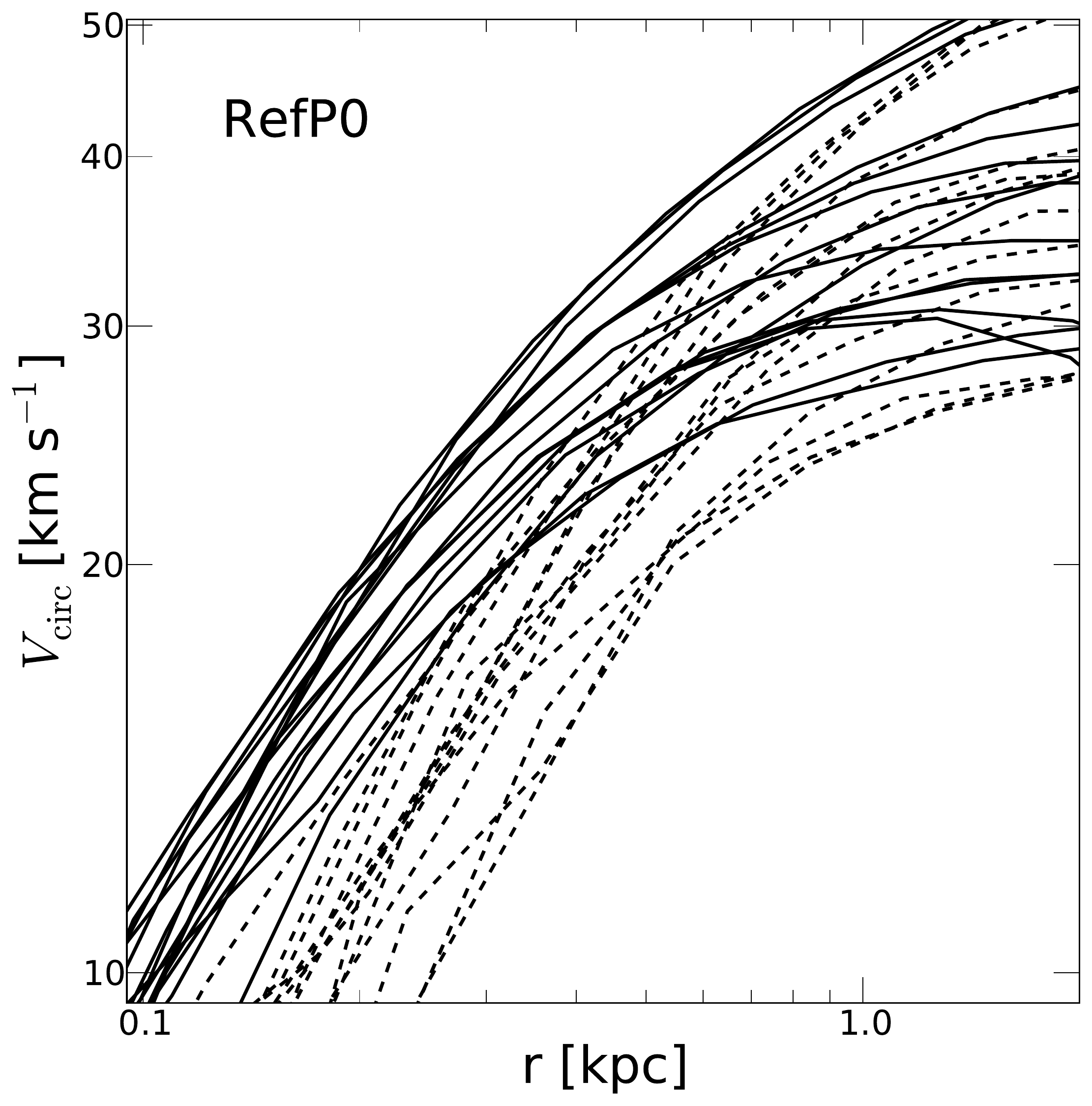}
\includegraphics[width=0.475\textwidth]{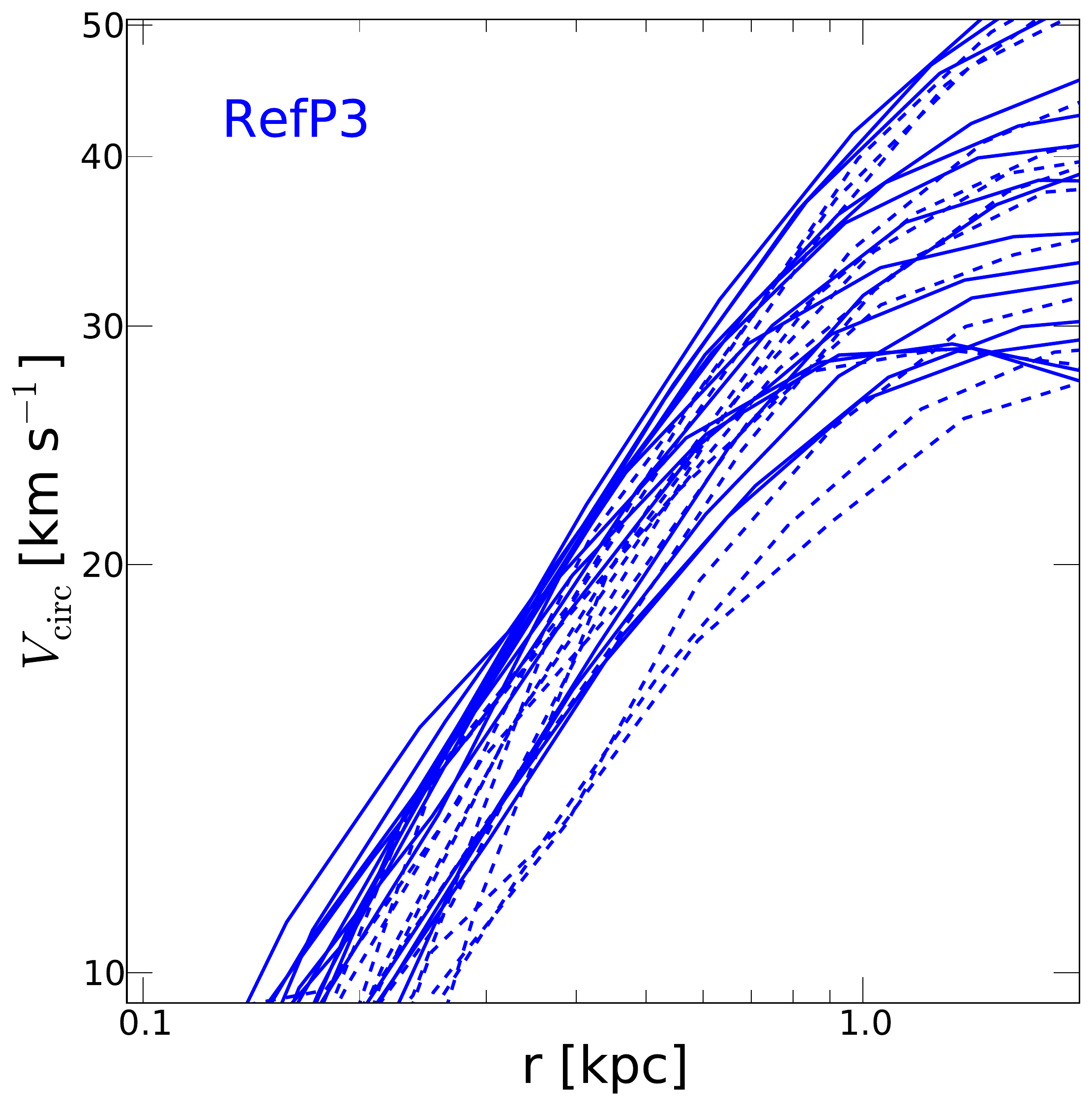}
\includegraphics[width=0.475\textwidth]{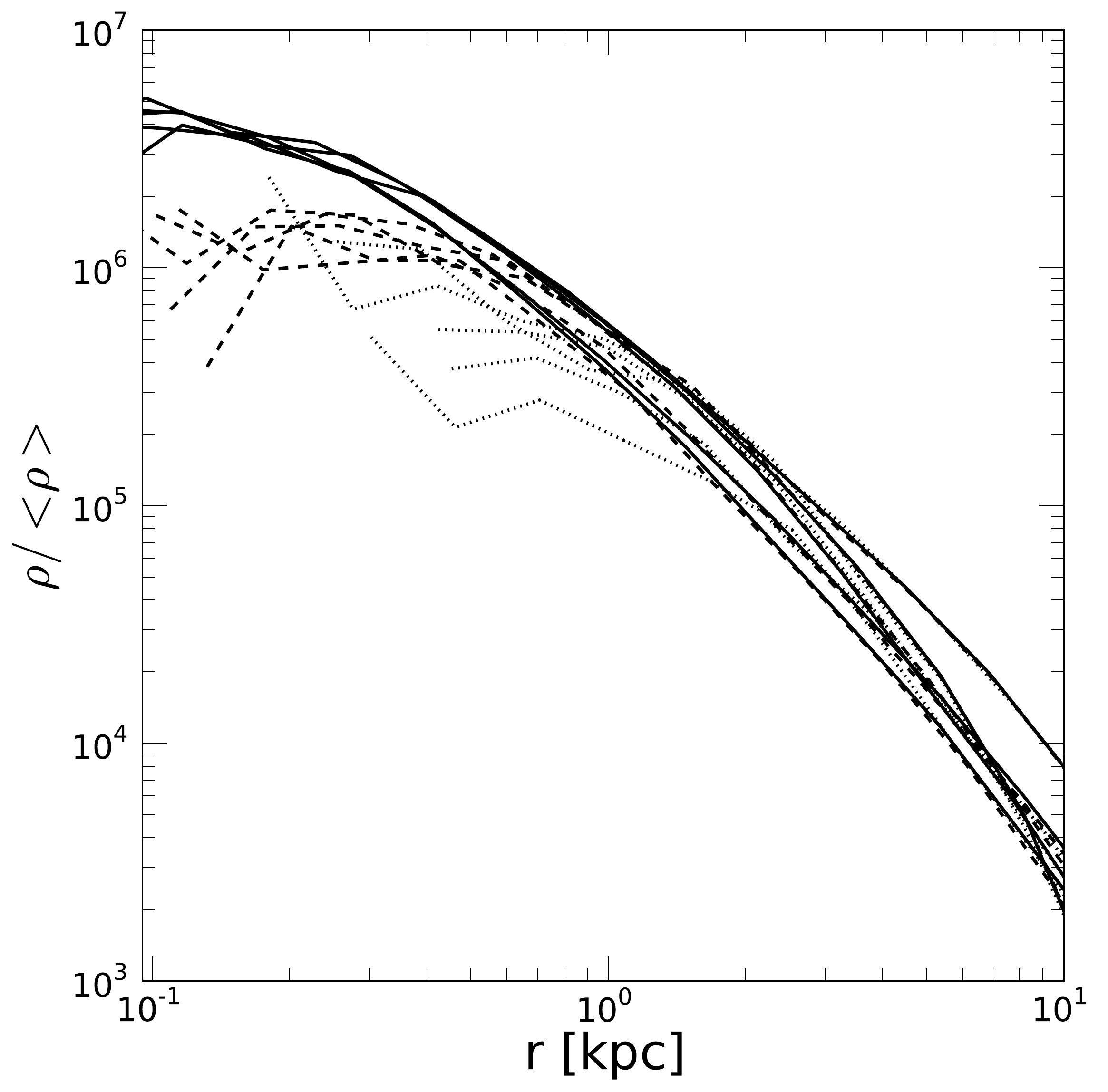}
\includegraphics[width=0.475\textwidth]{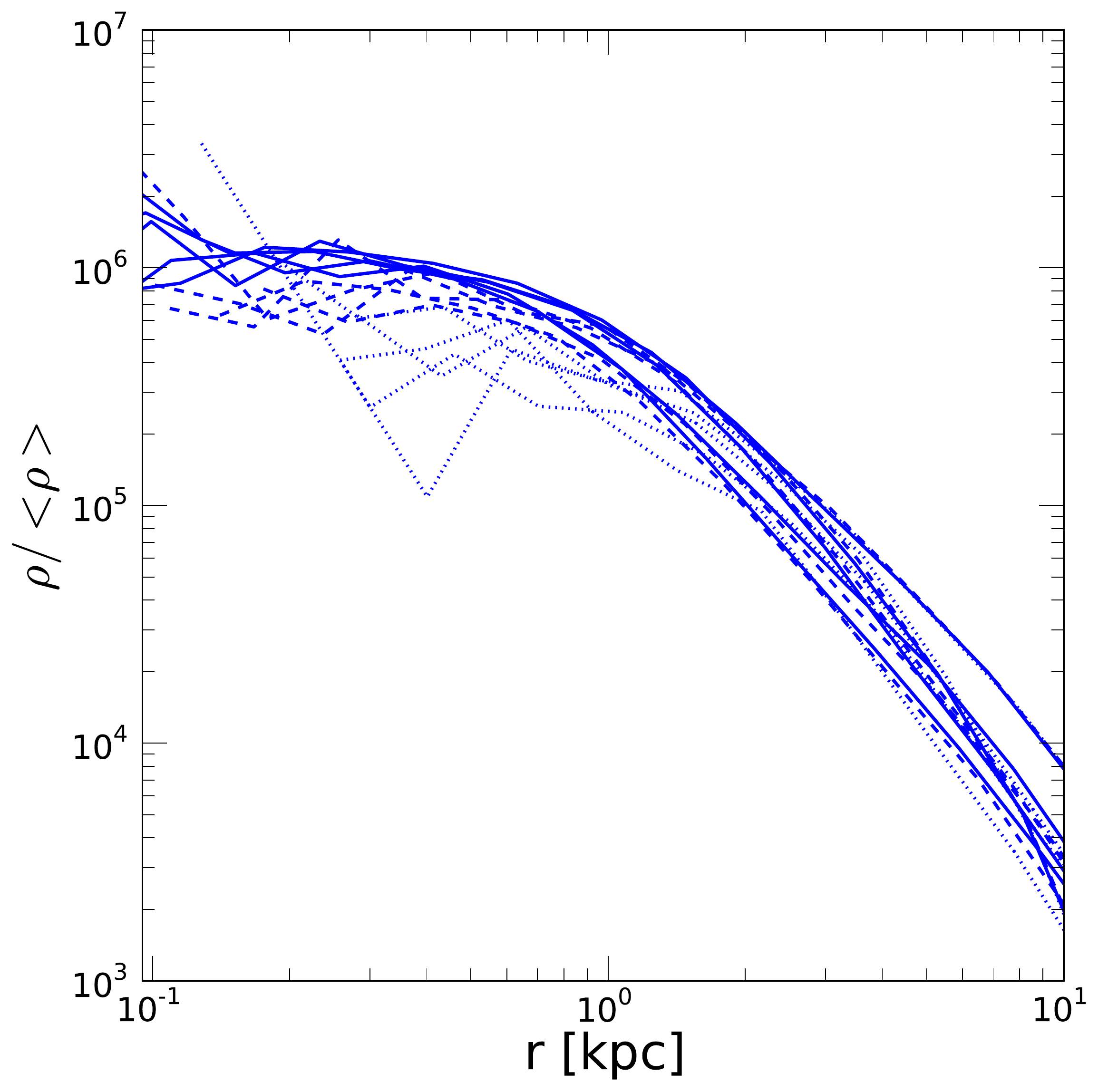}
\caption{Top panels: Circular velocity profiles for the $15$ most massive subhaloes (largest peak circular velocity) of the Aquarius-A halo
(left panel: RefP0, right panel: RefP3) at $z=0$
for two resolution levels: level 4 (dashed lines) and level 3 (solid lines). Note that the CDM subhalo
profiles are not converged yet, whereas those of the vdSIDM simulation show reasonable convergence due to the central core that builds
up in these cases. Although not shown here, this is also true for the RefP1 and RefP2 cases. Bottom panels: Density profiles for the $5$ most massive
subhaloes (left panel: RefP0, right panel: RefP3) at $z=0$ for three resolution levels: level 5 (dotted), level 4 (dashed lines) and level 3 (solid lines). CDM
profiles converge towards cuspy profiles, which are not fully resolved with level 3 resolution. The cored profiles of the vdSIDM simulation
shows reasonable convergence at level 3.}
\label{fig:sub_conv}
\end{figure*}

Since the MW dSphs are among the most DM-dominated galaxies, having very large 
dynamical mass-to-light ratios \citep[e.g.][]{Gilmore2007}, their stars are reliable dynamical tracers of their underlying DM subhaloes.
Since the available data for dSphs provides information only in three dimensions, whereas the phase-space distribution function is 6-dimensional,
the derived radial mass profiles from this data usually carries a degeneracy with the anisotropy profile which prevents model-independent
constrains on the underlying mass profile using a standard Jeans analysis (making for example a cored and a NFW profile equally acceptable by the data). 
Analyses that make use of more information in the data can extract more information about the mass profile.  This has been
accomplished not just with consideration of multiple stellar components, but also through Schwarzschild modelling \citep[][]{Jardel2011} and
indirectly based on arguments about survival of cold substructure within Ursa Minor \citep{Kleyna2003} and the distribution of globular clusters
in Fornax \citep{Goerdt2006}. Models that successfully reproduce the observed data tend to have roughly the same value of the enclosed mass within
the half-light radius \citep[e.g.][]{Strigari2007,Pena2008,Walker2009,Wolf2010}. In particular, \citet{Wolf2010} found that the value 
of the mass enclosed within the 3-dimensional de-projected half-light radius $r_{1/2}$ is accurately given by 
$M_{1/2}\approx3G^{-1}\left<\sigma_{\rm los}^2\right>r_{1/2}$ (or equivalently $V_{\rm circ}^2(r_{1/2})\approx3\left<\sigma_{\rm los}^2\right>$), 
where $\left<\sigma_{\rm los}^2\right>$ is the luminosity-weighted square of the line-of-sight velocity dispersion. This approximation is valid
as long as the velocity dispersion profile $\sigma_{\rm los}(R)$ remains relatively flat out to the 2-dimensional projected half-light radius. 
The quality of the kinematic data for the sample of dSphs we have mentioned, guarantees the estimation of $M_{1/2}$ with high accuracy.

In Figure~\ref{fig:sub_vcirc} we show the circular velocity profiles at $z=0$ of the 15 most massive subhaloes 
(largest peak circular velocity) in our simulations for  RefP0-3 from top left to bottom right. The black symbols with error 
bars represent the estimates of the circular velocity at the half-light radius for the 9 MW dSphs of the sample described above (the data was taken from Table I of \citealt{Wolf2010}). For the SIDM models (RefP1-3) we show our highest resolution results (level 3),
but for RefP0 the circular velocity curves are taken from the level 2 Aq-A-2 simulation ($2.8 \times 65.8 \sim 184$~pc spatial resolution)
since the level 3 simulation is not converged in that case due to the cuspyness of the subhalo density profiles. 
Convergence is easier to achieve for the SIDM models, because they do not develop cuspy profiles, but have constant
density cores as shown in Figure~\ref{fig:sub_density}. This convergence is explicitly demonstrated in Figure~\ref{fig:sub_conv} (top panels) where we
show the circular velocity curves of the $15$ most massive subhaloes for RefP0 (left panel) and RefP3 (right panel) at two levels of resolution:
level 4 (dashed lines) and level 3 (solid lines). Clearly vdSIDM subhaloes have essentially converged circular velocity profiles,
whereas CDM subhaloes are still moving towards a more concentrated mass distribution with increasing resolution\footnote{Although we do not show the RefP1 and RefP2 cases
in Figure~\ref{fig:sub_conv}, they also show good convergence as the RefP3 case.}. The bottom panels of Figure~\ref{fig:sub_conv} show the density profiles of the five
most massive subhaloes at all three resolutions (level 5 as dotted lines). The cores are clearly visible in the vdSIDM simulation along with the better convergence. 
We therefore conclude that the velocity profiles of the
SIDM simulations shown in Figure~\ref{fig:sub_vcirc} are not significantly affected by numerical effects. As for the RefP1 curves in Figure~\ref{fig:sub_density} 
we note one of the shown subhaloes has already entered the core-collapse regime by z=0, which is also clearly visible from the circular velocity curves.

As is evident from Figure~\ref{fig:sub_vcirc}, the vdSIDM models, which do not violate any of the astrophysical constraints, 
create a subhalo population whose most massive systems are dynamically consistent with the 
bright dSps. Although the subhalo cores are not as large as in the case of a constant self-scattering cross section (RefP1, top right), 
they are substantial enough to alleviate the tension with the data present in the CDM case (RefP0). We emphasise that although the most massive 
subhaloes have essentially the same maximum rotational velocities, i.e. the same mass, in the vdSIDM cases as in CDM, 
the vdSIDM subhaloes are much less concentrated and are therefore consistent with the observational data. 

Looking at Figure~\ref{fig:sub_vcirc} we see that at least three of the dSphs (Fornax, Sextans and Canes Venatici I, that have a value of
$r_{1/2}\sim1$~kpc) are inconsistent with the circular velocity profile of the 15 most massive subhaloes in all our SIDM simulations, 
particularly for the cases with a velocity-dependent cross section. This however, does not mean that there are no subhaloes in these simulations
that are a good match to these galaxies. These subhaloes exist but are considerably less massive than the ones we show in Figure~\ref{fig:sub_vcirc}
and are more affected by the limited resolution of our simulations. We note that even in the CDM case there seems to be enough subhaloes 
that are consistent with these three galaxies, although a possible concern is the high number of massive subhaloes that can be potentially 
compared with the dSphs. This is already evident in the CDM case: looking at the top leftmost panel of Figure 3 of \citet{Boylan2011b} (corresponding to
the Aq-A halo), we see that in order to find subhaloes that are consistent with all the 9 MW dSphs, it is necessary to consider approximately 30 subhaloes.
This issue is to be expected for the vdSIDM cases we have explored here as well. We believe however that this is a problem that is more likely related
to the stochastic nature of the abundance of the subhalo population in the Aquarius haloes. For instance, looking at the second panel from the top left in
Figure 3 of \citet{Boylan2011b}, it seems that Aq-B (another of the MW-like haloes simulated in the Aquarius project) has approximately 15
subhaloes that can be compared with the dSphs, which would clearly eliminate the aforementioned problem. Of course, a more detailed analysis needs to 
be done to firmly conclude this. Such analysis would include, in addition to possibly re-simulating other Aquarius haloes, making a one-to-one 
matching of the simulated subhalo population with the observed dSphs in a similar way to
what was done in \citealt{Boylan2011b}. In that work, those subhaloes that are considered Magellanic Cloud analogs are removed from the sample, 
and also the subhalo ranking is done according to their mass at infall (just before they merged with the host halo), which is a better
proxy of the potential well that shaped the luminous galaxy than the subhalo mass at $z=0$. We emphasise that the objective of this paper is
to present an initial exploration of the vdSIDM models discussed in Section \ref{mod_sec} and show explicitly that contrary to the CDM case,
these vdSIDM models do not predict a number of subhaloes which are too concentrated to host the bright dSphs.

\section{Summary and Conclusions}\label{conclusion}

The observed abundance and properties of dwarf galaxies have been an enduring challenge for
the remarkably successful CDM paradigm for more than a decade now. Recent observations with
high quality kinematic data seem to confirm that at least some of the DM-dominated MW dSphs have central
density cores instead of the high density cusps predicted by CDM \citep{Walker2011}. 
At the same time, the actual number of observed dwarf galaxies in the field (as inferred from the HI ALFALFA survey) seems to be in tension with 
the large abundance of dwarf haloes predicted by the CDM model \citep{ALFALFA2011,Ferrero2011}. This
seems to be a more serious challenge than the well known excess of DM subhaloes compared to the
number of observed MW satellites, pointed out over a decade ago \citep{Klypin1999,Moore1999},
and that has a viable solution based on the quenching of star formation \citep[e.g.][]{Koposov2009}.
Another potential ``failure'' of the CDM model has been raised very recently by \citet{Boylan2011a}, who
showed that the subhaloes kinematically associated to the MW dSphs seem to be much less concentrated than the most massive
subhaloes of current CDM simulations. A solution to all these challenges based exclusively on processes related to the 
formation an evolution of the luminous galaxies within the dwarf DM haloes can not completely be ruled out currently.
For instance, there are suggestions to create cores in CDM haloes as a response of the cuspy profile to energetic feedback.
Recent hydrodynamical simulations seem to be successful in producing cores if strong feedback mechanisms are invoked 
\citep[e.g.][]{Maccio2011,Pontzen2011}. It is however important to consider viable alternatives to the CDM model that preserve its success
on large scales and that, at the same time, are able to solve the small-scale problems. 

One of these alternatives is to consider the possibility that DM is collisional (SIDM first proposed by 
\citealt{SpergelSteinhardt2000}), with self-interactions that are able to isotropise the orbits of DM particles
and create a core in the inner regions of the haloes. In order for this alternative to be feasible, the DM-DM scattering cross section 
should avoid current astrophysical constraints, particularly stringent in clusters of galaxies that are consistent with the CDM model
\citep[for a summary of constraints see][]{Buckley2010}, and at the same time be large enough to
produce large density cores in dwarf galaxies.

In this paper we make a first assessment of the viability of a class of velocity-dependent SIDM (vdSIDM) models \citep{LoebWeiner2011} using high resolution numerical
simulations. More specifically we explore a simplified particle physics model where the self-scattering between DM particles is set by 
an attractive Yukawa potential mediated by a new gauge boson (either scalar or vector) in the dark sector \citep{Feng2010b, Finkbeiner2011,LoebWeiner2011}. 
This model, naturally predicts a transfer cross section that depends on velocity in such a way that can produce 
a significant effect in dwarf galaxies and avoid astrophysical constraints on larger scales (see Section \ref{mod_sec}). 

We implement an algorithm, based on a Monte Carlo approach, to account for DM self-interactions within the framework of this SIDM model in the 
GADGET-3 code for cosmological simulations; the description and testing of this algorithm can be found in Section \ref{tech}.
We use this modified code to run simulations of a MW-like halo using the initial conditions of one of the haloes of the Aquarius project 
\citep[Aq-A,][]{Springel2008} exploring two reference points within the allowed parameter space of the SIDM model for a elastic scattering case
(RefP2 and RefP3; see Table \ref{table:ref_points} and section \ref{IC}). 
As a reference, we also explored a constant cross section SIDM model (RefP1) which is ruled out by observations in galaxy clusters (see Sections \ref{mod_sec} and \ref{IC}). We find that for both of these cases, 
the density profile of the main halo remains the same as in the standard CDM case outside $\sim1$~kpc, whereas inside this radius a very small core 
develops (see Section \ref{main_halo} and Figure~\ref{fig:mainhalo_density}). The subhalo abundance and radial number density distribution are
not affected in the vdSIDM models we explored (Figure~\ref{fig:sub_stat}), which implies that in the elastic case, these models have the same
problem as the CDM model regarding the excess of low-mass subhaloes. On the other hand, the internal density profile (within $\sim1$~kpc) of the most massive
subhaloes change significantly due to the formation of a constant density core in their centre (Figure~\ref{fig:sub_density}). 
We show that the circular velocity profiles of these subhaloes are consistent with the kinematics of the brightest MW dSphs (Figure~\ref{fig:sub_vcirc}).
Specifically, and contrary to the CDM case, there are no subhaloes that are more concentrated than what is inferred from the kinematics of 
these galaxies. We therefore conclude that the SIDM models explored here significantly reduce the tension encountered between the 
observational data and the standard CDM predictions. 

The highest numerical resolution of the simulations presented in this work ($m_p=4.911\times10^4$~M$_{\odot}$, $\epsilon=120.5$~pc) is sufficient to obtain converged density and velocity profiles
for the more massive subhaloes (Figure~\ref{fig:sub_conv}), but is not enough to reliable trust those for smaller mass subhaloes. In a future work, we 
plan to make a more detailed analysis of the subhalo population extending to these small-mass systems and use an abundance matching technique (based
on the mass of the subhaloes at infall, which is better correlated with the luminosity of the galaxy) to make a one-to-one matching between the 
simulated subhaloes and the MW dSphs. We expect that this
will further reduce the tension between the dark subhalo population and the observed dSphs, and strengthen the case for vdSIDM models.
An additional improvement of our algorithm is to explore the possibility of inelastic scattering. In particular, in models where excited, nearly-degenerate,
DM states are possible, exothermic collisions could ``evaporate'' subhaloes by ejecting particles during down-scatterings. This could potentially alleviate 
the problem on the overabundance of low mass subhaloes as well.

\section*{Acknowledgements}

The simulations in this paper were run on the Odyssey cluster supported by the
FAS Science Division Research Computing Group at Harvard University.
We thank Volker Springel for giving us access to {\sm GADGET-3} and Tracy Slatyer and Douglas Finkbeiner
for useful discussions. We also thank Matthew Walker, Simon D.M. White and the referee Naoki Yoshida for useful comments
that improved the paper. JZ is supported by the University of Waterloo and
the Perimeter Institute for Theoretical Physics. Research
at Perimeter Institute is supported by the Government of
Canada through Industry Canada and by the Province of
Ontario through the Ministry of Research $\&$ Innovation. JZ
acknowledges financial support by a CITA National Fellowship.
This work was supported in part by NSF grant AST-0907890 and NASA grants
NNX08AL43G and NNA09DB30A (for A.L.)

\label{lastpage}

\end{document}